\documentclass{jpp}

\usepackage{graphicx}
\usepackage{natbib}

\ifCUPmtlplainloaded \else
  \checkfont{eurm10}
  \iffontfound
    \IfFileExists{upmath.sty}
      {\typeout{^^JFound AMS Euler Roman fonts on the system,
                   using the 'upmath' package.^^J}%
       \usepackage{upmath}}
      {\typeout{^^JFound AMS Euler Roman fonts on the system, but you
                   dont seem to have the}%
       \typeout{'upmath' package installed. JPP.cls can take advantage
                 of these fonts, if you use 'upmath' package.^^J}%
      }
  \else
  \fi
\fi

\ifCUPmtlplainloaded \else
  \checkfont{msam10}
  \iffontfound
    \IfFileExists{amssymb.sty}
      {\typeout{^^JFound AMS Symbol fonts on the system, using the
                'amssymb' package.^^J}%
       \usepackage{amssymb}%
         \let\leq=\leqslant
         
      }{}
  \fi
\fi

\ifCUPmtlplainloaded \else
  \IfFileExists{amsbsy.sty}
    {\typeout{^^JFound the 'amsbsy' package on the system, using it.^^J}%
     \usepackage{amsbsy}}
    {}
\fi

\newsavebox{\astrutbox}
\sbox{\astrutbox}{\rule[-5pt]{0pt}{20pt}}

\title[Physics of radio emission in gamma-ray pulsars]{Physics of radio emission in gamma-ray pulsars}

\author[S. A. Petrova]
{S. A. Petrova
  \thanks{Email address for correspondence: petrova@rian.kharkov.ua}}

\affiliation{Institute of Radio Astronomy, NAS of Ukraine,
Chervonopraporna Str., 4, Kharkov 61002, Ukraine}

\pubyear{2016}
\volume{}
\pagerange{--}
\date{?; revised ?; accepted ?. - To be entered by editorial office}
\begin{document}

\maketitle

\begin{abstract}
Propagation of radio emission in pulsar magnetosphere is reviewed. The effects of polarization transfer, induced scattering and reprocessing to high energies are analysed with an especial emphasis on the implications for the gamma-ray pulsars. The possibilities of the pulsar plasma diagnostics based on the observed radio pulse characteristics are outlined as well. As an example, the plasma number density profiles obtained from the polarization data for the Vela and the gamma-ray millisecond pulsars J1446-4701, J1939+2134 and J1744-1134 are presented. The number densities derived tend to be the highest/lowest when the radio pulse leads/lags the gamma-ray peak. In the PSR J1939+2134, the plasma density profiles for the main pulse and interpulse appear to fit exactly the same curve, testifying to the origin of both radio components above the same magnetic pole and their propagation through the same plasma flow in opposite directions. The millisecond radio pulse components exhibiting flat position angle curves are suggested to result from the induced scattering of the main pulse by the same particles that generate gamma-rays. This is believed to underlie the wide-sense radio/gamma-ray correlation in the millisecond pulsars. The radio quietness of young gamma-ray pulsars is attributed to resonant absorption, whereas the radio loudness to the radio beam escape through the periphery of the open field line tube.
\end{abstract}

%

\section{Introduction}
Pulsar radio emission is generally associated with the processes in the secondary electron-positron plasma produced in the polar gap and outflowing along the open magnetic field lines \citep[for a review of pulsar radio emission mechanisms see, e.g.,][]{l95,m03,l08}. As the radio emission originates deep in the open field line tube, it should propagate through the plasma flow. The propagation effects should substantially modify the radio pulse characteristics and prove to have pronounced observational manifestations. Furthermore, the propagation concept has a great potential for diagnostics of the pulsar plasma.

For the first 40 years of pulsar research, the picture of a pulsar was chiefly based on the comprehensive observational data on its radio emission. The numerous gamma-ray pulsar detections by Fermi LAT \citep{fc1,fc2} have convincingly demonstrated the incompleteness of the classical picture of a pulsar. As the gamma-ray emission with the observed spectral characteristics cannot originate in the polar gap \citep{fc1}, the actual pulsar model needs to include the regions of particle acceleration and very high energy emission in the outer magnetosphere or beyond the light cylinder. Moreover, the gamma-ray observations in conjunction with the already known or follow-up radio data have revealed the principally new features of the pulsar radio emission \citep{fc1,fc2}. In particular, the class of young energetic radio quiet pulsars has been established. As for the young radio loud pulsars, their radio pulses typically precede in phase the high-energy profiles. In contrast, in the gamma-ray millisecond pulsars (all of them are radio loud), the radio pulse components can be located differently with respect to the gamma-ray profile: they can either lead or lag the main gamma-ray peak or roughly coincide with it.

These results are traditionally interpreted on geometrical grounds. As the radio beam originates deep in the open field line tube, it is much narrower than the high-energy emission pattern and, correspondingly, can much more frequently miss an unfavourably located observer, making the phenomenon of a radio quiet gamma-ray pulsar quite abundant. The gamma-ray light curves, including their location relative to the radio profile, are modelled extensively for a number of assumptions as to the gamma-ray emission region. The gamma rays are associated with the outer, slot and partially screened polar gap \citep{v14,pier15}, with the acceleration region of the finite-conductivity plasma beyond the light cylinder of the force-free magnetosphere \citep{k14,b15}, and with the magnetospheric current sheet \citep{u14,spitk15}. Each of the models fits the observational data satisfactorily, and at the moment neither of them can be ruled out.

As a rule, the gamma-ray pulsar studies exploit the radio pulse data only in the geometrical aspect. It should be noted, however, that the pulsar radio emission bears a wealth of valuable information. In the present paper, we concentrate on the physics of pulsar radio emission, placing particular emphasis on the case of gamma-ray pulsars. Sect.~2 contains the basics of the radio wave propagation theory. We outline the observational consequences of the propagation effects and demonstrate their potential to probe the pulsar plasma. Sect.~3 is devoted to interpretation of radio peculiarities of the gamma-ray pulsars in terms of the propagation effects. It is shown that the radio quietness of young pulsars can be attributed to significant synchrotron absorption. The radio pulse components of millisecond pulsars, which lag the main high-energy peak or roughly coincide with it, are suggested to result from induced scattering of the pulsar radio beam in the outer magnetosphere. In Sect.~4, the problems and prospects are discussed.

\section{Propagation effects in pulsar plasma}  
\subsection{Polarization transfer}
The ultrarelativistic highly magnetized plasma of pulsars allows two non-damping natural waves, the ordinary and extraordinary ones \citep[e.g][]{ms77,ab86,lyut98a,m99}. In case of negligible gyrotropy, the two types of waves are polarized linearly, with the electric vector lying in the plane of the ambient magnetic field and perpendicularly to this plane, respectively. The two-stream instability, which is most probably responsible for the pulsar radio emission in the inner magnetosphere \citep{i70,s70,uu88,a90,w94,am98,lp00}, cannot generate the extraordinary waves directly, because of their vacuum dispersion \citep[e.g.][]{l95}. The non-damping ordinary waves are superluminal and can arise as a result of induced scattering of the subluminal ones off the plasma particles in the radio emission region \citep{l96}.

In a number of pulsars, the radio emission is indeed characterized by a high percentage of linear polarization, with the position angle swing across the pulse reflecting the change of the ambient magnetic field projection onto the plane of the sky as the pulsar rotates \citep{rc69}. In general, however, the empirical picture of radio pulsar polarization appears much more complicated. As a rule, pulsar radio emission is a mixture of the two polarized modes, which generally have elliptical polarization and can be markedly non-orthogonal. The ellipticities and position angles of the modes as well as their intensity ratio fluctuate strongly from pulse to pulse.

The observed polarization peculiarities can be attributed to polarization transfer of radio waves in pulsar magnetosphere. Moreover, propagation origin of the radio pulse polarization implies a unique possibility to probe the pulsar plasma based on the observed polarization profiles.

In the vicinity of the emission region, the characteristic scale length for beating between the natural modes, $L_b\sim c/(\omega\Delta n)$ (where $\omega$ is the wave frequency, $\Delta n$ the difference of the mode refractive indices, $c$ the speed of light), is much less than the scale length for change in the plasma characteristics, $L_p\sim r$ (where $r$ is the altitude above the neutron star). Therefore the geometrical optics approximation is valid, in which case the electric vector of the modes follows the local orientation of the ambient magnetic field \citep{cr79,b86}. Pronounced polarization evolution can be expected in the regions where $\Delta n\to 0$ and the regime of geometrical optics is broken.

\subsubsection{Linear conversion}
The refractive indices of the natural waves become almost equal in the region of quasi-longitudinal propagation with respect to the ambient magnetic field. In general, ray propagation in the open field line tube is quasi-transverse, $\theta\gg 1/\gamma$ (where $\theta$ is the wavevector tilt to the magnetic field, $\gamma$ the plasma Lorentz-factor). However, due to refraction of radio waves in pulsar plasma, in the inner magnetosphere the ray trajectory may contain a short segment almost parallel to the magnetic field. It is the place where the natural waves suffer linear conversion \citep{p01}. The original ordinary waves become an incoherent mixture of the ordinary and extraordinary ones, which further on propagate independently. The total intensity is conserved, and the mode intensity ratio is determined by the characteristics of the plasma in the region of conversion. The anticorrelation of polarization mode intensities found in PSR B0329+54 \citep{e04} strongly suggests that it is the linear conversion that underlies the phenomenon of orthogonally polarized modes in pulsar emission. Being interpreted in terms of linear conversion, the mode intensities and their longitudinal and pulse-to-pulse fluctuations can be regarded as a probe of the pulsar plasma distribution and its temporal variations.

\subsubsection{Polarization-limiting effect}
As the open field line tube widens with distance from the neutron star, the plasma number density decreases and the refractive index of the ordinary waves tends to unity. Correspondingly, in the outer magnetosphere, the geometrical optics approximation is again broken and the waves are subject to the polarization-limiting effect \citep{cr79,b86,lp98a,pl00}. In the polarization-limiting region, the electric vector of the waves has no time to follow the ambient magnetic field direction and wave mode coupling starts. As a result, each of the incident natural waves becomes a coherent mixture of the two natural modes of the ambient plasma, acquiring some circular polarization and position angle shift. As the two types of waves evolve identically, the outgoing radiation presents a superposition of the two purely orthogonal elliptically polarized modes. This is in line with the empirical model of pulsar polarization \citep[e.g.][]{ms00}.

\subsubsection{Diagnostics of pulsar plasma}
The ellipticity and position angle shift resulting from the wave mode coupling are related to each other, both being determined by the plasma characteristics in the polarization-limiting region. Based on this relation, a technique of number density diagnostics of pulsar plasma was developed \citep{p03a}. It allows to obtain the plasma number density distribution in the open field line tube proceeding from the observed polarization radio profile of a pulsar.

At a fixed altitude $r_0$ in the tube, the number density $N_0$ changes as function of the polar angle $\chi$. Besides that, because of continuity, the plasma flow widens with altitude together with the open field line tube in such a way that
\[
N\left (r,\frac{\chi}{\chi_f(r)}\right )=N_0\left (\frac{\chi}{\chi_f(r_0)}\right )\left (\frac{r}{r_0}\right )^{-3},
\]
where $N$ is the number density at the altitude $r$ and $\chi_f$ is the polar angle of the tube boundary at the corresponding altitude. Furthermore, it is convenient to normalize $N$ to the Goldreich-Julian number density \citep{gj69},
\[
\kappa=\frac{NPce}{B},
\]
where $B$ is the magnetic field strength at the same altitude $r$, $P$ the pulsar period, $c$ the speed of light, and $e$ the electron charge. The resultant multiplicity $\kappa$ is independent of altitude and generally characterizes the number of secondary particles per primary particle (the theories of the polar gap pair creation cascade yield the multiplicuty values in the range from less than unity up to $\sim 10^5$ \citep{ha01b,ae02,ml10,timhar15}; in the outer gap scenario, the multiplicities may also be as high as $\sim 10^4-10^5$ \citep[e.g.,][]{twc10}; in the wind zone, even higher values, $\sim 10^6-10^8$, are expected \citep[e.g.,][]{tt13}). Thus, the plasma number density profile presented in terms of $\kappa(\chi/\chi_f)$ is a universal characteristics of the plasma distribution inside the open field line tube of a pulsar. It should be noted that since the radio beam coming into the polarization-limiting region usually occupies a small part of the tube cross-section, only a part of the plasma density distribution can be reconstructed from the observed polarization profile. Making use of the multifrequency polarization data allows to extend the region available for the plasma diagnostics substantially. 

\begin{figure}
\vspace{5cm}
  \centerline{\includegraphics[height=8cm,width=8cm]{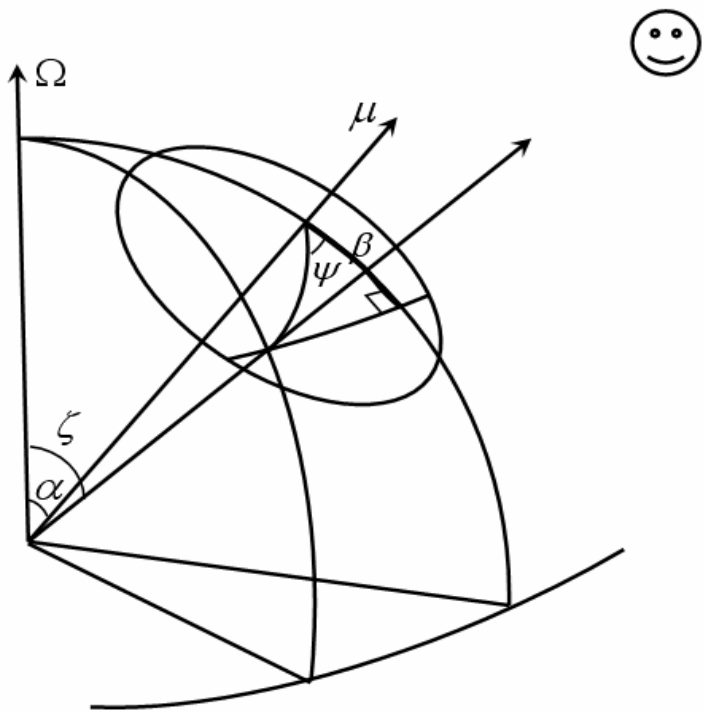}}
    \caption{Pulsar geometry; the magnetic moment and the line of sight are inclined to the rotation axis at the angles $\alpha$ and $\zeta$, respectively, $\beta\equiv\zeta-\alpha$ is the impact angle, and $\psi$ the geometrically defined position angle.}
\label{f0}
\end{figure}

\begin{figure}
  \centerline{\includegraphics[height=8cm,width=12cm]{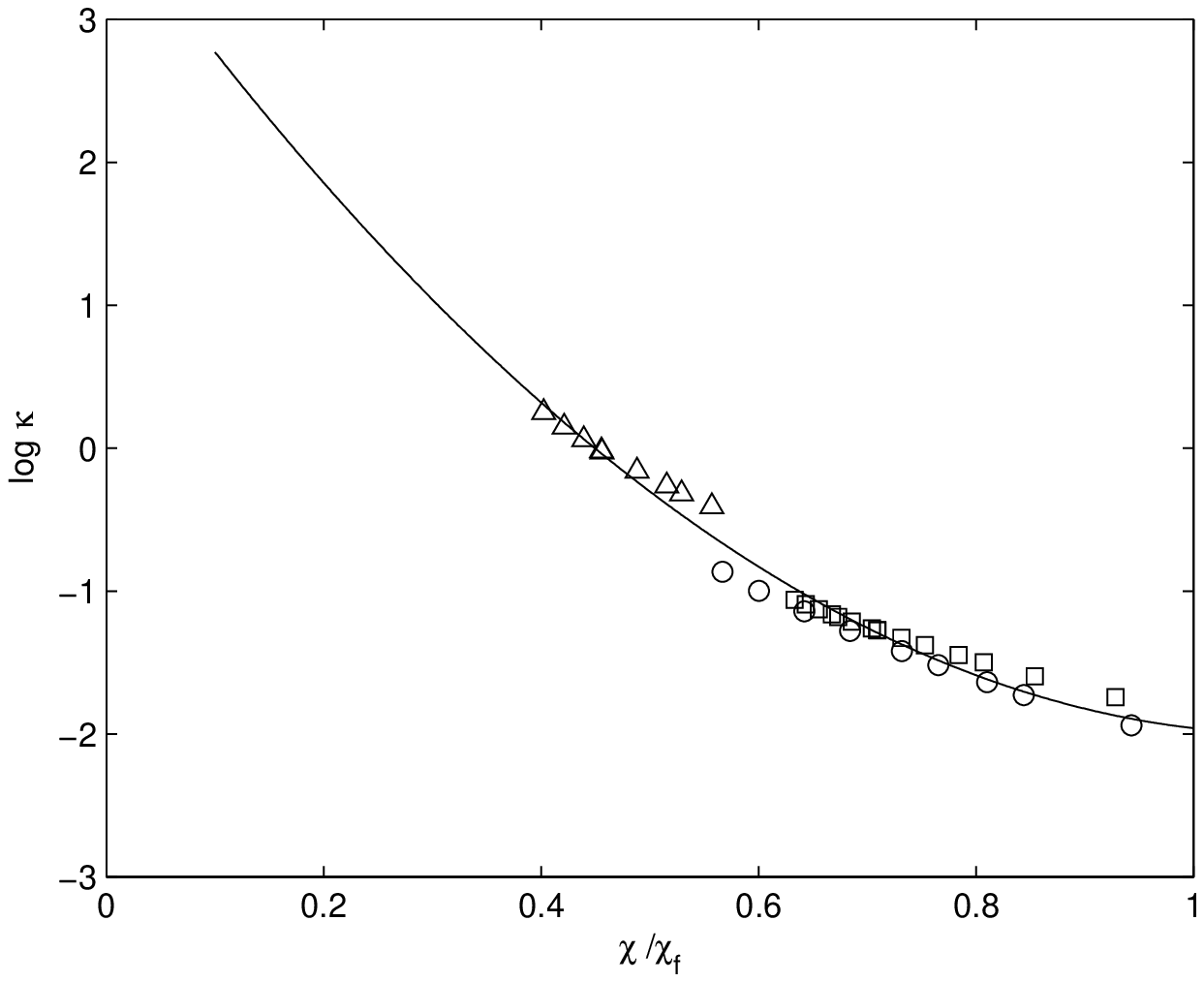}}
  \centerline{\includegraphics[height=8cm,width=12cm]{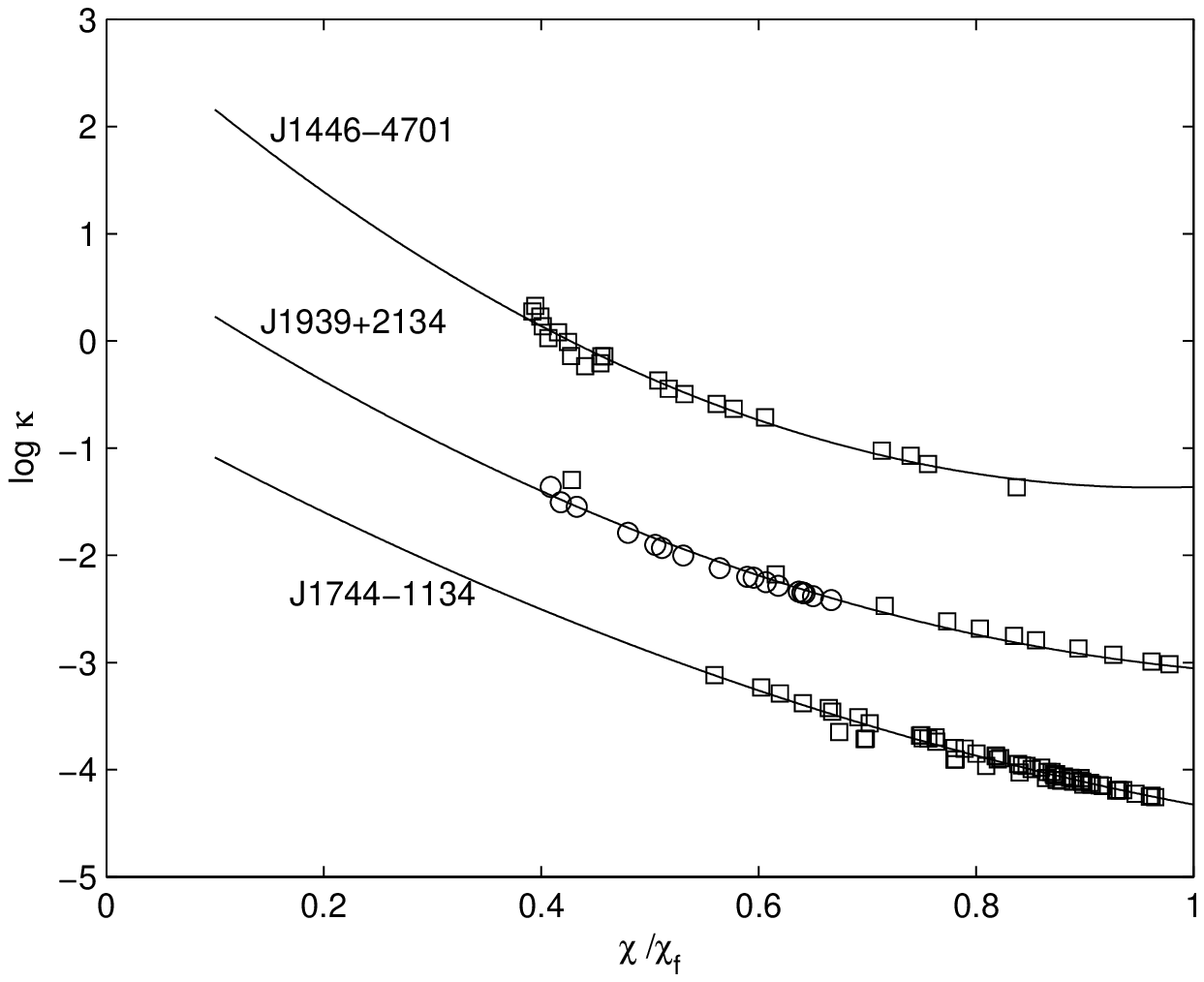}}
  \caption{Plasma density profiles for the Vela (\textit{upper panel}) and the millisecond  gamma-ray pulsars J1446-4701, J1744-1134, and J1939+2134 (\textit{lower panel}); $\kappa$ is the plasma multiplicity defined as the number density normalized to the Goldreich-Julian number density, $\chi$ the polar angle of a point inside the open field line tube, $\chi_f$ the polar angle of the tube boundary at the corresponding altitude. The circles, squares and triangles in the upper panel mark the points obtained from the polarization data of \citet{vela1,vela2} at 1.4, 3.1 and 8.4 GHz, respectively. The points in the lower panel are obtained from the polarization profiles of \citet{dai15} at 1.4 GHz; the squares and circles correspond to the main pulse and interpulse data, respectively. The polarization profiles are available via the EPN Database of Pulsar Profiles at http://www.epta.eu.org/epndb. The geometry assumed is as follows: $\alpha=35.6^\circ$, $\beta=4^\circ$ for the Vela, $\alpha=60^\circ$, $\beta=5^\circ$ for J1446-4701, $\alpha=40^\circ$, $\beta=1.5^\circ$ for J1744-1134, and $\alpha=79^\circ$, $\beta=4^\circ$ for J1939+2134, where $\alpha$ is the pulsar inclination and $\beta$ the impact angle.}
\label{f1}
\end{figure}

Figure~\ref{f0} illustrates the relevant geometry. Figure~\ref{f1} exhibits examples of application of the technique to the average polarization radio profiles of several gamma-ray pulsars. The resultant plasma density profiles show exponential decrease towards the edge of the open field line tube, with the exponent being well fitted by a second-order polynomial. (Such a character of the plasma density distribution is similar to that found for the longer-period pulsars PSR B0355+54 and PSR B0628-28 in \citep{p03a}).

The class of young energetic gamma-ray pulsars is presented by the Vela pulsar (see Fig.~\ref{f1}, upper panel). As can be seen, the points resulting from the polarization data at different frequencies well match the same curve, confirming the validity of the technique used. The plasma multiplicity values appear in line with the results of numerical simulations of pair cascades in the polar gap \citep{ha01b,ae02,ml10}, though they depend appreciably on the pulsar geometry assumed.

The lower panel in Fig.~\ref{f1} shows the plasma density profiles for three gamma-ray millisecond pulsars, PSRs J1446-4701, J1939+2134 and J1744-1134, which are the representatives of the three classes, with the radio pulse components leading, coincident with and lagging the main gamma-ray peak \citep[cf.][]{fc2}. As is evident from the figure, the plasma multiplicities for the three pulsars differ substantially, being the highest for the leading and the lowest for the trailing position of the radio pulse. In the former case, the plasma multiplicities are more or less similar to those of the Vela pulsar, where the radio profile also leads the high-energy peak. The latter is in a general agreement with the interpretation in terms of the partially screened gap \citep[e.g.][]{v14}, in which case the open field line tube is deficient of pair plasma.

Our result, however, should be taken with caution, since the scaling of the plasma density profiles along both axes noticeably depends upon the poorly known pulsar geometry. To the best of our knowledge, the only constraints on the geometrical parameters of the pulsars considered are obtained from gamma-ray light-curve modeling \citep[e.g.][]{v14}. Keeping in mind these results, we have taken the values which provide $\chi/\chi_f\leq 1$ for the points of the plasma density profile. In fact, the values $\chi/\chi_f>1$ may be quite realistic, since for the millisecond pulsars the assumption of a purely dipolar magnetic field structure involved in our technique can hardly be regarded as appropriate. Note that the plasma density profiles appear to lie within $\chi/\chi_f\leq 1$ for quite special geometry, whereas over a broad range of geometrical parameters $\chi/\chi_f$ may be a few times larger, in which case the multiplicities are typically much higher as well.

The only advantage of our choice of the pulsar geometry and of our approach on the whole is definiteness. At present, the force-free model \citep[e.g.,][]{spitk06,abin15}, its dissipative versions \citep[e.g.,][]{k12,lst12} and MHD-extension \citep[e.g.,][]{3d13} provide a more realistic representation of the pulsar magnetosphere. These models, being much more advanced as compared to the classical vacuum dipole model, contain a number of additional ingredients, such as the global electric current, the accelerating electric field, etc. One can expect that incorporating all this into the theory of polarization transfer would complicate it substantially, and the transparent unambiguous relation between the observed quantities, which underlies the above discussed technique of the plasma diagnostics, would no longer be the case. It is yet to be understood what are the main features of the polarization evolution of radio waves in the force-free magnetosphere and what it may imply for the plasma diagnostics. It should also be kept in mind that, since the particle motion in the force-free fields is principally distinct from the motion of a probe particle in the fixed fields of the same strengths, the polarization transfer in the force-free magnetosphere should be considered anew, e.g., based on the self-consistent two-fluid model \citep{br00,ko09,p15}.

One more important point can be noticed in the lower panel of Fig.~\ref{f1}. For the pulsar PSR J1939+2134, the points of the plasma density profile obtained from the main pulse and interpulse polarization data follow exactly the same curve. (Note that the uncertainty of the pulsar geometry discussed above does not affect the validity of this result). Given that the main pulse and interpulse originate on both sides of the neutron star, such an exact match would seem improbable. This improbable coincidence disappears if the main pulse and interpulse were to come from the same region, as in the bidirectional model of pulsar radio emission \citep{d05,p08a}, where the two components are emitted from the same pole in the outward and backward directions, respectively (for details on the physical grounds of the model see Sect. 2.2.2 below). General considerations as to the origin of the radio components coincident with high-energy peaks and lagging them are presented in Sect. 3.2. As was pointed out by the Referee, the bi-directional model for the gamma-ray emission might be physically realizable in the current sheet inside the light cylinder, in which case the space-like current should be supplied by the counter-streaming particle beams.

The technique of plasma density diagnostics developed in \citet{p03a} and discussed above is based on a simplistic assumption of a cold non-gyrotropic plasma embedded in a superstrong dipolar magnetic field. In this case, the ellipticity and position angle shift of outgoing waves exhibit one-to-one correspondence, both being determined by the plasma characteristics in the polarization-limiting region. The position angle shift resulting from polarization evolution in the gyrotropic plasma is suggested as a potential probe of the global current of a pulsar \citep{ha01a}. Further generalizations of the theory of polarization transfer in pulsar plasma by including realistic energy distributions of the plasma particles as well as finiteness and non-dipolar character of the magnetic field \citep{p06a,p06b,wang10,bp12} lead to a complicated picture of observational consequences. Because of its numerous ingredients, the picture is no longer transparent and can hardly allow to solve an inverse problem of plasma diagnostics. Nevertheless, the elaborated theory can explain a number of observed polarization peculiarities and seems appropriate for direct modeling of the pulsar polarization profiles.

\subsubsection{The role of cyclotron absorption in polarization evolution}
In the vicinity of the radio emission region, the magnetic field is so strong that the electron gyrofrequency is much higher than the radio wave frequency in the particle rest frame, $\omega_H\equiv eB/mc\gg \omega^\prime$. As the magnetic field strength decreases with distance from the neutron star, in the outer magnetosphere the radio wave frequencies meet the cyclotron resonance condition, $\omega^\prime=\omega_H$. As a rule, the cyclotron resonance occurs above the polarization-limiting region. If the two regions lie close to each other, the plasma number density in the resonance region is still high enough to cause further evolution of radio wave polarization \citep{p06a}. The cyclotron absorption coefficients of the ordinary and extraordinary natural modes are slightly different: the extraordinary waves are suppressed somewhat stronger. The waves coming into the resonance region are already a coherent mixture of the two natural modes, and since these components are absorbed differently, the outgoing polarization changes. In this case, the evolution of the two orthogonally polarized modes is distinct and they finally become non-orthogonal.

If the waves pass through the resonance before the polarization-limiting region, the extraordinary ones are strongly suppressed, whereas the ordinary ones suffer weaker polarization evolution, with the resultant circular polarization having the opposite sense \citep{p06a}.

The pulse-to-pulse variations of the radio pulse polarization can be attributed to the temporal fluctuations of the plasma characteristics. In particular, the fine peculiarities of single-pulse polarization statistics observed in PSR B0329+54 \citep{es04} and PSR B0818-13 \citep{e04} can be reproduced by numerical modeling of polarization transfer in the plasma with weak pulse-to-pulse fluctuations of the number density \citep{p06b}.

\subsection{Induced scattering}
The brightness temperatures of pulsar radio emission, $T_B\sim 10^{25}-10^{30}$ K, are so high that spontaneous scattering off the plasma particles is strongly dominated by the induced one \citep{bs76,wr78}. For typical conditions in the pulsar magnetosphere, the latter process can be efficient in a number of regimes. Induced scattering of plasma waves is thought of as an important constituent of the pulsar radio emission mechanism \citep{l96}. Besides that, it can affect further propagation of radio waves in the magnetosphere.

Within the framework of the plasma mechanisms, pulsar radio emission is believed to be generated at the local Lorentz-shifted proper plasma frequency, $\omega\sim\omega_p\sqrt{\gamma}$ (for the criticism of this statement see \citet{mg99}). Therefore close to the emission region the induced scattering is a collective process. The transverse waves may also participate in the induced three-wave processes \citep{lm06}. A particular case of induced Raman scattering in pulsar magnetosphere was examined in \citet{gang93,lyut98b}.

As the plasma number density rapidly decreases with distance from the neutron star, far enough from the emission region the collective effects are negligible. Then the plasma can be considered as a system of independent particles, whereas the incident radiation presents a mixture of the transverse electromagnetic waves polarized in the plane of the ambient magnetic field (the ordinary mode) and perpendicularly to this plane (the extraordinary mode).

In the open field line tube of a pulsar, induced scattering by a system of particles may hold in different regimes, which are distinct in the role of the ambient magnetic field, in the transverse momentum of the scattering particles and in the characteristics of the final state of the scattered radiation. The induced scattering may lead to a substantial redistribution of radio intensity in pulse longitude and frequency and may have multitudinous observational consequences. Note that the characteristics of the radiation scattered in different regimes can be regarded as a tool for diagnostics of pulsar plasma in different regions of the magnetosphere.

\subsubsection{The case of superstrong magnetic field}
In the inner magnetosphere, where the number density of the scattering particles is the largest, the induced scattering is expected to be most efficient. In this region, the radio frequency in the particle rest frame is much less than the electron gyrofrequency, $\omega^\prime\equiv\omega\gamma(1-\beta\cos\theta)\ll\omega_H$ (where $\omega$ and $\theta$ are the photon frequency and tilt to the ambient magnetic field in the laboratory frame, $\beta$ and $\gamma$ are the particle velocity in units of $c$ and Lorentz-factor). Then the magnetic field affects the scattering process, changing both the scattering cross-section and the geometry of particle recoil \citep{c71,bs76}. In the regime of superstrong magnetic field, the scattering particles slide along the magnetic field lines and the transverse component of their perturbed motion in the field of an incident wave is suppressed as well. Correspondingly, only the photons with the ordinary polarization are involved in this type of scattering.

As the pulsar radio emission is related to the processes in the ultrarelativistic secondary plasma, in each point the radiation is concentrated into a narrow beam with the opening angle $\sim 1/\gamma$. However, the beams of substantially differing frequencies may have different orientations as a result of refraction and ray propagation in the rotating magnetosphere. The problem of induced scattering of radio photons between the two beams of substantially different frequencies, $\omega_1$ and $\omega_2$, and orientations with respect to the ambient magnetic field, $\theta_1$ and $\theta_2$, is considered in \citet{p04b}. Of course, in the rest frame of the scattering particles the beam frequencies are equal, $\omega_1^\prime=\omega_2^\prime$, and hence, in the laboratory frame they are related as $\omega_1(1-\beta\cos\theta_1)=\omega_2(1-\beta\cos\theta_2)$.

In case of a superstrong magnetic field, the induced scattering between the two beams of intensities $I_1$ and $I_2$ is shown to result in the net intensity transfer from the lower-frequency beam to the higher-frequency one, with the total intensity of the two beams remaining approximately constant, $I_1+I_2=\mathrm{Const}$. For the decreasing spectrum of pulsar radio emission, $I\propto\omega^{-\xi}$ (where $I$ is the original intensity of the pulsar radiation and $\xi$ the spectral index) this may imply significant amplification of the higher-frequency intensity. For example, if $\theta_1/\theta_2\approx 3$, then $\omega_2/\omega_1\approx 10$ and for $\xi=3$ the original intensity ratio of the beams is $I_1/I_2=(\omega_1/\omega_2)^{-\xi}\sim 10^3$. Correspondingly, in case of efficient scattering the higher-frequency intensity may increase by three orders of magnitude. Thus, the induced scattering can account for the observed energetics of giant pulses.

It is evident that the amplification is stronger for the pulsars with steeper radio spectra. Steepness of radio spectrum was suggested as a criterion for the search of new giant pulse sources \citep{p04b}. The three sources discovered subsequently do meet this criterion \citep{k06a,k06b}.

Propagation origin of giant pulses implies that their intensity statistics are determined by pulse-to-pulse fluctuations in the pulsar plasma. Given that the plasma number density and the original radio intensity are Gaussian random quantities, the distribution of strongly amplified pulses has a power-law form and may fit the intensity distributions of the observed giant pulses \citep{p04b}.

Induced scattering inside the narrow beam leads to photon focusing in the direction closest to the ambient magnetic field \citep{p04a}. As a result, the beam width may decrease by about two orders of magnitude, becoming compatible with the characteristic scale of the observed microstructure. In giant pulses, the focusing effect is believed to be much stronger and can be responsible for their substructure at nanosecond timescales. Note that the focusing effect resolves the difficulty of too high brightness temperatures of the nanopulses observed in the Crab pulsar \citep{h03}.

Generally speaking, the power-law statistics and extremely short-scale structures are characteristic of non-linear processes. In particular, the power-law intensity distribution of giant pulses and their nanosecond substructure can also be interpreted in terms of the wave collapse resulting from the nonlinear modulation instability \citep{w98,c04}.

\subsubsection{Induced scattering into background}
Because of strong directivity of pulsar radio emission, the photon direction corresponding to the maximum scattering probability lies outside the photon beam. Hence, induced scattering into this direction is possible only in the presence of background radiation. The latter may arise, e.g., as a result of spontaneous scattering of the beam photons. Although the intensity of the background radiation is negligible as compared to the beam intensity, $\log I_\mathrm{bg}/I_\mathrm{beam}\approx -10$, in case of strong enough induced scattering, with the scattering depth $\Gamma\approx 30$, a substantial part of the original beam intensity can be transferred into the background, the photons being predominantly scattered in the direction of the maximum scattering probability \citep{lp96}.

In the regime of superstrong magnetic field, the scattered component is directed approximately along the scattering particle velocity and the induced scattering is most efficient at the upper boundary of the scattering region \citep{p08c}. Because of rotational aberration in that region, in the pulse profile observed the scattered component should lead the main pulse by $\leq 30^\circ$ and therefore can be identified with the precursor. Within the framework of this mechanism, the complete linear polarization of the precursor emission and its spectral peculiarities are explained naturally.

In the regime of moderately strong magnetic field, ($1/\sqrt{\gamma}\ll\omega^\prime/\omega_H\ll 1$), the induced scattering of pulsar radio emission into background can also be efficient \citep{p08a}. In this case, the perturbed motion of the scattering particle presents a drift in the crossed electric field of the incident wave and the ambient magnetic field \citep{bs76,ochusov83}. Correspondingly, the scattering geometry and the properties of the scattered radiation are essentially distinct. In particular, the photons of both polarizations are involved in the scattering, whereas the scattered component is antiparallel to the particle velocity and appears in the pulse profile as an interpulse. A comprehensive picture of the observed interpulse properties may well be explained in terms of induced scattering \citep{p08a}.

It is important to note that the scattering nature of the precursor and the interpulse implies a physical relation between these components and the main pulse. The observational manifestations of such a relation are already known. The first example concerns the pulsar PSR B1822-09 \citep{f81,g94}. In the weak emission mode its radio profile exhibits the main pulse and interpulse, whereas in the bright mode it contains the precursor, the main pulse and much weaker interpulse. Thus, the precursor and interpulse are anticorrelated, and their appearance is controlled by the main pulse intensity. The second example is the pulsar PSR B1702-19 \citep{welt07}, in which case the subpulse structure of the main pulse and interpulse is modulated with exactly the same periodicity, testifying to the correlation of the instantaneous intensities of the two components.

\subsubsection{Induced scattering by relativistically gyrating particles}
The radio emission structure of the Crab pulsar presents an exceptional case. The radio profile contains a total of seven components spread out over the whole pulse period and characterized by distinct spectral and polarization properties \citep{mh96,mh99}. At the lowest frequencies, the profile consists of the precursor, the main pulse and the interpulse, which have very steep spectra and disappear at higher frequencies. As the frequency increases, the precursor is replaced by the so-called low-frequency component (LFC) located at the earlier pulse phase, and instead of the interpulse there appear the high-frequency interpulse (IP$^\prime$) also shifted toward earlier phases as well as the pair of trailing high-frequency components (HFC1 and HFC2).

In order to explain this complicated picture, the theory of induced scattering is generalized to the case of relativistically gyrating particles \citep{p08b,p08d,p09b}. In the inner magnetosphere, the magnetic field of a pulsar is so strong that any transverse momentum of the plasma particles is almost immediately lost via synchrotron re-emission. At higher altitudes, however, the particles acquire relativistic gyration as a result of resonant absorption of radio emission (see Sect. 2.3 below). As the pulsar radio beam is broadband, $\nu\sim 10^7-10^{10}$ Hz, the resonance region appears quite extended in altitude. Therefore the lower-frequency radio waves are subject to strong-field scattering by the particles performing relativistic gyration because of resonance absorption of the higher-frequency radio emission.

At the conditions relevant to pulsar magnetosphere, the induced scattering by relativistically gyrating particles may be efficient for several lowest harmonics of the gyrofrequency. In particular, the precursor and LFC of the Crab pulsar can be attributed to the main pulse scattering at the zeroth harmonic and from the first to the zeroth one, respectively \citep{p08d}. In this case, the location of the two components on the pulse profile, their spectral behaviour and high linear polarization are explained naturally.

Furthermore, the interpulse, IP$^\prime$, HFC1 and HFC2 can be interpreted as the consequence of subsequent induced scattering of the main pulse, precursor and LFC, respectively, in the regime of moderately strong magnetic field \citep{p09b}. Then HFC1 and HFC2 are actually a single component split by rotational aberration close to the light cylinder. Note that approximate constancy of the position angle of linear polarization across this component \citep{mh99} does testify to its origin in the outer magnetosphere.

As is shown in \citet{p09b}, the fine structure in the observed dynamic spectrum of giant IP$^\prime$s \citep{he07,h15} can be identified with the structure of the giant main pulses modified by induced scattering in the channel 'main pulse $\to$ precursor $\to$ IP$^\prime$'. The fine structure of the giant main pulses itself is recently interpreted in terms of induced scattering as well \citep{t15}.

The implications of induced scattering of pulsar radio emission are not exhausted by those outlined above. The scattered components of the millisecond pulsars are addressed in Sect. 3.2.

\subsection{Radio photon reprocessing to high energies}
The radio emission, whose luminosity is only a tiny fraction of the total luminosity of a pulsar, is believed to be generated by some coherent mechanism acting in the plasma flow above the polar gap \citep[e.g.,][]{l95,m03,l08}. The pulsar high-energy emission is commonly associated with the incoherent curvature, synchrotron and inverse Compton emission from the slot \citep[e.g.,][]{mh04,hard08} or outer \citep{romani96,cheng00,hirot06} gap or from the current sheet both inside and outside the light cylinder \citep{spitk15}. In case of very-high-energy emission, the mechanism is either inverse Compton emission from the outer magnetosphere \citep{lyut12,lyut13,hk15} or synchrotron emission from the region beyond the light cylinder \citep{petri12,u14,k14,mp15,b15}.

Thus, pulsar emissions in the radio and higher-energy ranges are essentially distinct. Nevertheless, their correlation is strongly expected, since any emission is undoubtedly attributed to the plasma particles. Any observational manifestation of such a correlation is believed to be a unique probe of the pulsar plasma characteristics and the pulsar magnetosphere on the whole. For the two decades, a targeted search for correlation between the giant radio pulses and gamma-ray emission of the Crab pulsar has failed to yield positive results \citep{lund95,rt98,bilous11,bilous12,aliu12}. Instead, the observational manifestations of the radio -- high-energy connection were found at softer energies. (For the discussion of the radio/gamma-ray correlation in the millisecond pulsars see Sect. 3.2.)

The optical pulses coincident with giant radio pulses of the Crab pulsar are 3\% brighter than the average pulse \citep{shearer03, oosterbr08,s13}. Although the Vela pulsar does not exhibit classical giant pulses, it is known for its prominent intensity variations accompanied by radio profile changes \citep{kd83}. Namely, the intensity ratio of the precursor and the main pulse increases with radio intensity. Furthermore, the X-ray profile of the Vela pulsar integrated over the range 2-16 keV is found to change with radio intensity as well \citep{lommen07}. The changes touch the main peak and the leading trough, which presents a separate component at softer energies and passes through the spectral peak at a few tenths keV \citep{hard02}.

Note that in both pulsars the high-energy fluctuations correlated with radio emission are much less marked than the radio pulse changes and, in the Vela case, touch not only the total flux but also the profile shape. The connection of such a type can hardly be attributed to the global changes in the magnetosphere and can rather be interpreted in terms of radio photon reprocessing to high energies.

\subsubsection{Resonant absorption of radio emission}
In the inner magnetosphere, the plasma particles are at the ground Landau orbital due to extremely strong synchrotron re-emission. In the outer magnetosphere, in the region of cyclotron resonance of the radio waves, the particles can absorb and emit resonant photons, performing transitions between the Landau orbitals. In total, the incident radio emission is partially absorbed, whereas the particle pitch-angle $\psi\equiv p_\perp/p_\Vert$ (where $p_\perp$ and $p_\Vert$ are the transverse and longitudinal components of the particle momentum, respectively) increases monotonely. Note that because of much weaker magnetic field synchrotron re-emission no longer prevents the momentum growth. At the conditions relevant to pulsar magnetospheres the process of resonant absorption can be very efficient \citep{bs76,m82,lp98b}. On the one hand, this poses the problem of radio wave escape from the pulsar magnetosphere. On the other hand, as soon as the particles acquire substantial pitch-angles, the very character of absorption changes \citep{lp98b}.

At the initial stage, the particle pitch-angle grows rapidly up to the values $\psi\approx\theta\sim 0.1$ (where $\theta$ is the incident angle of the resonant photons). Later on the total particle momentum $p$ increases so as to satisfy the condition $p(\theta-\psi)={\rm Const}$. The estimates show that in pulsars only the highest radio frequencies, $\nu >10$ GHz, suffer resonant absorption in the regime $\psi\ll\theta$, whereas over the rest part of the radio range the radiation is absorbed in the regime $\theta -\psi\ll\theta$, in which case the particle momenta can increase by 2-3 orders of magnitude \citep{p02}.

The optical depth to resonant absorption in the regime $\theta -\psi\ll\theta$ is much less than that in the regime $\psi\ll\theta$, and it is the former quantity that determines the total radio luminosity of a pulsar. For the long-period and millisecond pulsars, the optical depth is small, and their radio emission freely escapes from the magnetosphere. In the short-period normal pulsars, the optical depth is roughly of order unity, so that their radio emission may be substantially suppressed. As is argued in Sect. 3.1 below, the resonant absorption can account for the radio quietness of young gamma-ray pulsars.

\subsubsection{Spontaneous synchrotron re-emission}
The synchrotron radiation of the particles with the evolved momenta falls into the optical and soft X-ray ranges. Spontaneous re-emission of the particles taking part in the resonant absorption of radio waves was suggested as a mechanism of the pulsar high-energy emission \citep{lp98b,g01,p03b}. The model was subsequently elaborated by including the accelerating electric field and applied to the cases of partially screened polar gap and slot gap \citep{hum05,hard08,hk15}. It is important to note that the mechanism based on synchrotron re-emission for the first time implies the physical connection between the radio and high-energy emissions of a pulsar \citep{p03b}. In particular, the observed excess of optical emission during giant radio pulses of the Crab pulsar \citep{shearer03,oosterbr08,s13} can be understood as follows. As a giant pulse comes into the resonance region, the momentum evolution of the particles becomes more pronounced, and their synchrotron emission is stronger.

\subsubsection{Spontaneous scattering by relativistically gyrating particles} 
At the conditions relevant to pulsar magnetospheres, radio photons can also be reprocessed to high energies in another way. Over most part of the resonance  region there is a significant amount of the under-resonance photons, with frequencies $\omega^\prime\ll\omega_H$, which can be scattered off the relativistically gyrating particles. This process differs essentially from the common magnetized scattering by straightly moving particles \citep{p08b}. The under-resonance photons are chiefly scattered to high harmonics of the particle gyrofrequency, $\omega_{\rm sc}^\prime =\omega^\prime +s\omega_H$ with $s\sim\gamma_0^3$ (where $\gamma_0$ is the Lorentz-factor of the particle gyration), and the total scattering cross-section is much larger. Furthermore, the peak of the scattered radiation is markedly shifted in frequency beyond the synchrotron maximum. Correspondingly, although the total power scattered is always less than the synchrotron power, in the region beyond the synchrotron maximum the contribution of the scattered radiation can be substantial. 

As is shown in \citet{p09a}, the spontaneous scattering of under-resonance radio photons off the relativistically gyrating particles can explain the radio -- X-ray correlation observed in the Vela pulsar \citep{lommen07}.
The scattered power appears to be a very strong function of the particle gyration energy, $L_{\rm sc}\propto\gamma_0^6$, and the quantity $\gamma_0$ is determined by the radio intensity coming into the resonance region and causing the particle momentum evolution. Thus, it is the scattered power that is tightly connected with the radio emission characteristics, and this connection should be most prominent beyond the synchrotron maximum. Recall that in the Vela pulsar the trough component is indeed noticeably connected with the radio emission beyond its spectral maximum. The numerical estimates as applied to this pulsar give reasonable results. In the observer frame, the peaks of the synchrotron and scattered radiation, $\hbar\omega_{\rm syn}=0.2$ keV and $\hbar\omega_{\rm sc}=1.5$ keV, well agree with the spectral maximum of the trough component and the range of its pronounced correlation with radio, while the total synchrotron power, $L_{\rm syn}=10^{31}$ erg\,s$^{-1}$, is compatible with the observed luminosity of this component.

The radio beam is also subject to induced scattering off the gyrating particles. In contrast to the spontaneous scattering considered above, the induced scattering is most efficient between the states below the resonance, $\omega_{\rm sc}^\prime=\omega^\prime$, and can modify the radio profile shape (for details see Sect. 2.2 above). Note that the efficiency of induced scattering is also a function of the particle gyration energy $\gamma_0$. Hence, the variations of $\gamma_0$ should lead to joint fluctuations in the radio and high-energy emission, and the radio -- high-energy connection in the Vela pulsar can be understood as the interplay between the processes of spontaneous and induced scattering by the gyrating particles \citep{p09a}.

\section{Implications for gamma-ray pulsars}
The radio propagation effects considered in the previous section can explain a number of peculiarities observed in the radio emission of gamma-ray pulsars and can potentially give valuable information as to the magnetospheric plasma of these objects. Different effects can be combined to form a comprehensive physical picture of the pulsar magnetosphere. Here we discuss the specific observational manifestations of radio propagation effects, which could only be revealed by joint radio and gamma-ray studies of pulsars.

\subsection{Young energetic pulsars}
One of the unexpected results of Fermi LAT is the discovery of an abundant population of radio quiet pulsars. About a half of young gamma-ray pulsars are radio quiet, whereas all known millisecond pulsars are radio loud \citep[e.g.,][]{fc2}. The radio quietness is commonly attributed to unfavourable orientation of the pulsar with respect to an observer \citep[see, e.g.,][and references therein]{fc2}. Such an interpretation is in part motivated by the absence of marked differences in the gamma-ray emission of the radio quiet and radio loud pulsars. Here we argue that the radio quietness may be caused by the physical rather than geometrical effect.

\begin{figure}
  \centerline{\includegraphics[height=10cm,width=15cm]{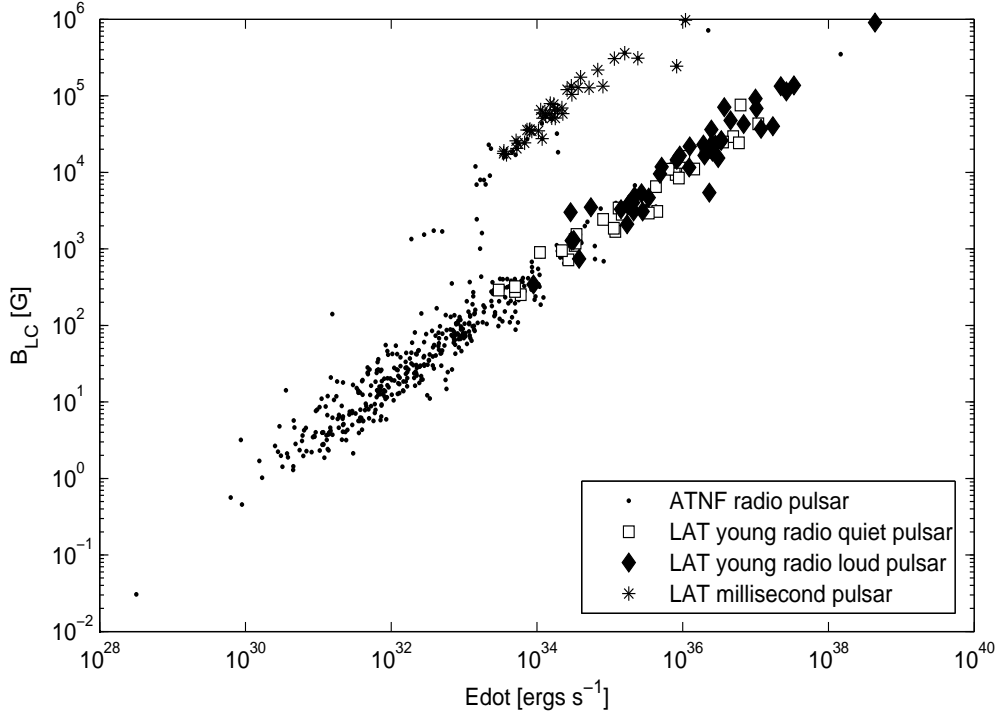}}
    \caption{Light-cylinder magnetic field vs. the rate of rotational energy loss. The radio pulsar data are taken from the ATNF Pulsar Catalogue (http://atnf.csiro.au/people/pulsar/psrcat), the gamma-ray pulsar data from the LAT Second Pulsar Catalog (http://fermi.gsfc.nasa.gov/ssc/data/access/lat/2nd\_PSR\_catalog).}
\label{f2}
\end{figure}

\begin{figure}
  \centerline{\includegraphics[height=7cm,width=10cm]{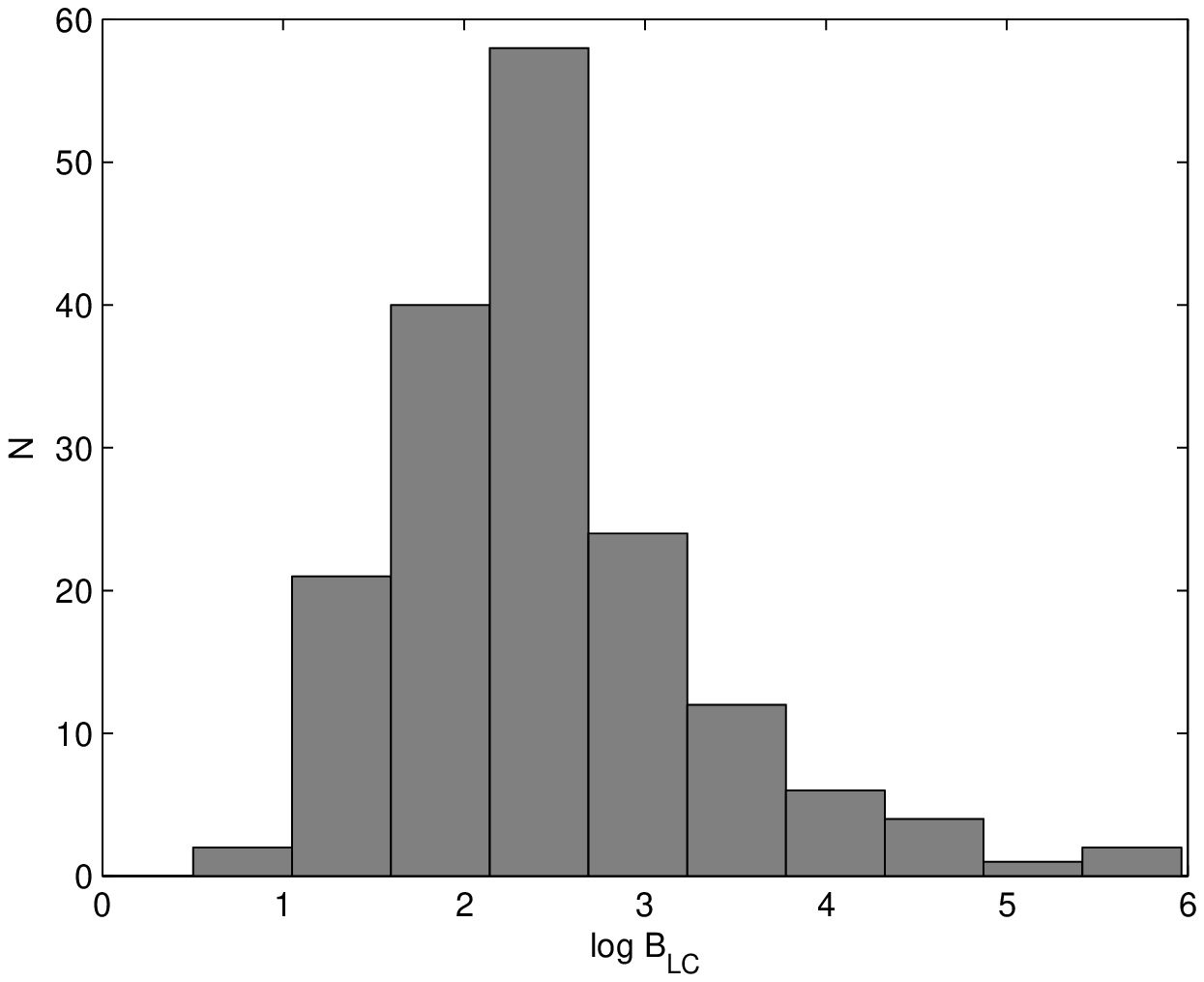}}
  \centerline{\includegraphics[height=7cm,width=10cm]{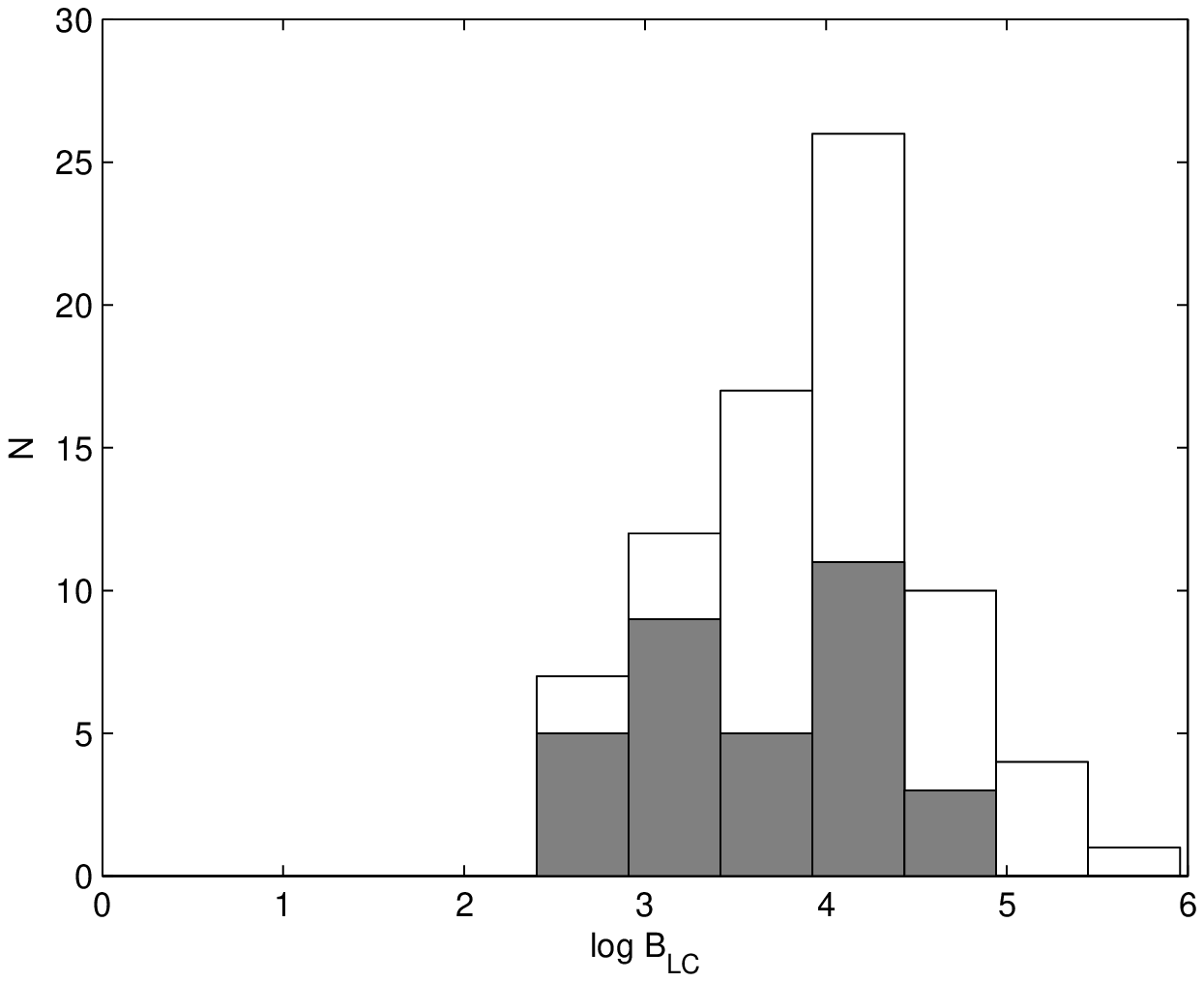}}
  \centerline{\includegraphics[height=7cm,width=10cm]{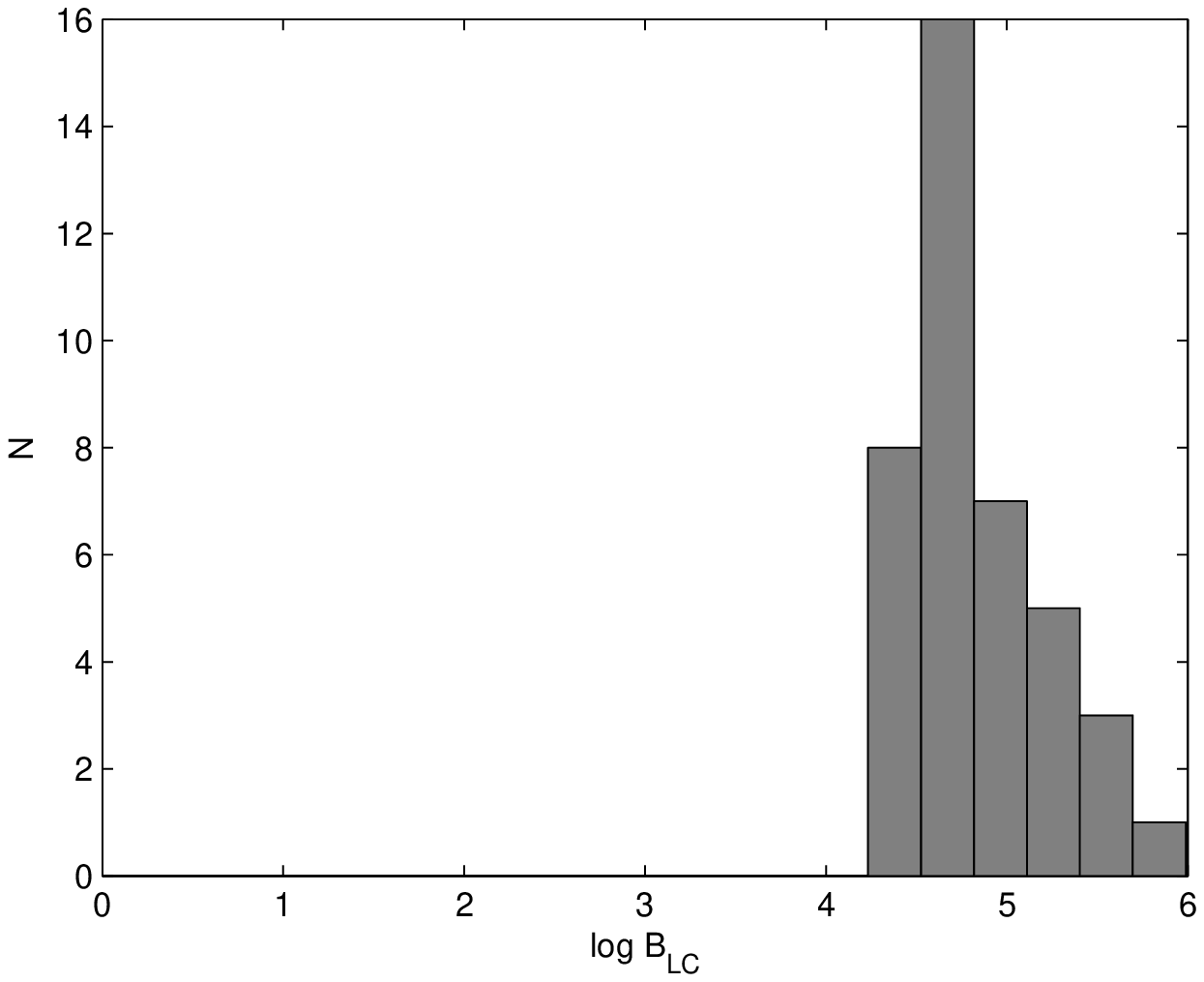}}
  \caption{Distribution of pulsars in the light-cylinder magnetic field; \textit{upper panel}: short-period ($0.03-0.5$ s) radio pulsars from the ATNF Pulsar Catalogue, \textit{middle panel}: young radio loud (white) and radio quiet (grey) gamma-ray pulsars from \citet{fc2}, \textit{lower panel}: gamma-ray millisecond pulsars from \citet{fc2}. }
\label{f3}
\end{figure}

As is demonstrated in Fig.~\ref{f2}, the light-cylinder magnetic field of the pulsars, $B_{\rm LC}$, increases with the rate of rotational energy loss, $\dot{E}$, with the young gamma-ray pulsars being characterized by the largest values of these quantities among the normal radio pulsars. The distribution of short-period ($P=0.03-0.5$ s) radio pulsars in $B_{\rm LC}$ (see Fig.~\ref{f3}, upper panel) peaks at $\log B_{\rm LC}\approx 2.5$ and shows noticeable asymmetry, testifying to the deficit of higher-$B_{\rm LC}$ pulsars. The population of young gamma-ray pulsars has $\log B_{\rm LC}> 2.5$, occupying the region beyond the peak (cf. Fig.~\ref{f3}, middle panel). Furthermore, as can be noticed from the latter figure, the radio quiet pulsars tend to have lower $B_{\rm LC}$ than the radio loud ones. Presumably this trend would be even more pronounced if the actual inclination angles and the realistic magnetic field geometry were taken into account.

The above results make us suggest that the radio quietness of pulsars can be explained in terms of resonant absorption. The problem of too high absorption depths in pulsars was recognized long ago \citep{bs76}. \citet{g01} attributed the radio quietness of Geminga to cyclotron absorption. The treatment of synchrotron absorption \citep{lp98b,p02} removed the problem of radio wave escape from the long-period and millisecond pulsars, whereas in the short-period normal pulsars the absorption depths can still be substantial (for details see Sect.~2.3.1).

Keeping in mind that the synchrotron absorption depth $\Gamma$ is a strong enough function of poorly known pulsar geometry, one can still expect that as long as $\log B_{\rm LC}< 2.5$ the value of $\Gamma$ is less than unity, allowing free escape of radio waves from the bulk of short-period normal radio pulsars. As $\Gamma$ increases with $B_{\rm LC}$, the pulsars corresponding to the region beyond the peak in the upper panel of Fig.~\ref{f3} may suffer substantial synchrotron absorption and become radio quiet. On the other hand, as $B_{\rm LC}$ increases further, the resonance region moves to higher altitudes in the magnetosphere. Then, given sufficiently rapid rotation of a pulsar, the radio beam meets the resonance at the periphery of the open field line tube. Thus, the radio waves acquire a chance to escape through the outer gap, the closed field line region, etc., making the pulsar radio loud. Note that the radio loud pulsars are indeed characterized by shorter periods \citep[e.g.,][]{fc2} and higher $B_{\rm LC}$ (see Fig.~\ref{f3}, middle panel).

As is found in \citet{marelli15}, the radio loud and radio quiet pulsars differ in the X-ray flux, though their X-ray spectra and gamma-ray properties are the same. Traditionally, the authors interpret this on geometrical grounds: they hypothesize on the existence of a separate X-ray component close to the magnetic axis, so that its visibility depends on the pulsar orientation with respect to an observer, similarly to the radio beam visibility.

The difference in the X-ray emission of the radio loud and radio quiet pulsars can also be interpreted in terms of propagation effects. Given that the radio flux is severely absorbed, the radio wave scattering by relativistically gyrating particles, which reprocesses the photons into the X-ray range (see Sect. 2.3.3) does not hold. Hence, the radio quiet pulsars lack the corresponding X-ray component. It is still questionable, however, if the radio photon reprocessing in radio loud pulsars can account for the energetics of the additional X-ray component observed.

\subsection{Millisecond pulsars}
The population of gamma-ray millisecond pulsars also exhibits increase of $B_{\rm LC}$ with $\dot{E}$ (see Fig.~\ref{f2}). The law of increase is roughly the same as in the young energetic pulsars, except for the overall shift toward lower $\dot{E}$ by about two orders of magnitude. Interestingly, the maximum $B_{\rm LC}$ and minimum $\dot{E}$ are the same for both populations, whereas the minimum $B_{\rm LC}$ in the millisecond pulsars roughly corresponds to the upper limit for the radio quiet pulsars (cf. the middle and lower panels in Fig.~\ref{f3}). This correspondence, along with a sharp cut-off in the distribution of the millisecond pulsars at low $B_{\rm LC}$ (see the lower panel in Fig.~\ref{f3}) excites a doubt as to the absence of radio quiet or radio dim millisecond pulsars in nature.

\begin{figure}
  \centerline{\includegraphics[height=10cm,width=15cm]{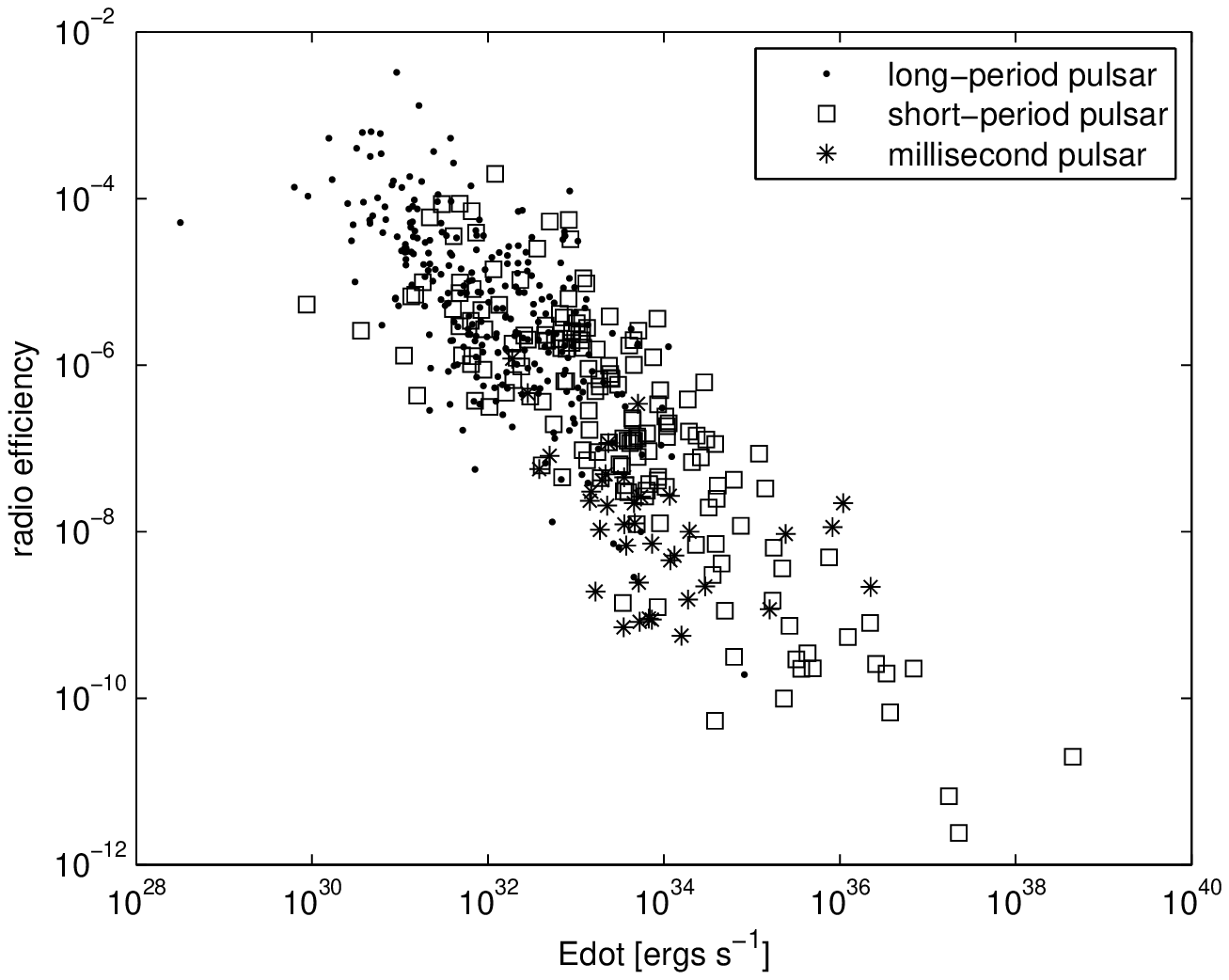}}
  \centerline{\includegraphics[height=10cm,width=15cm]{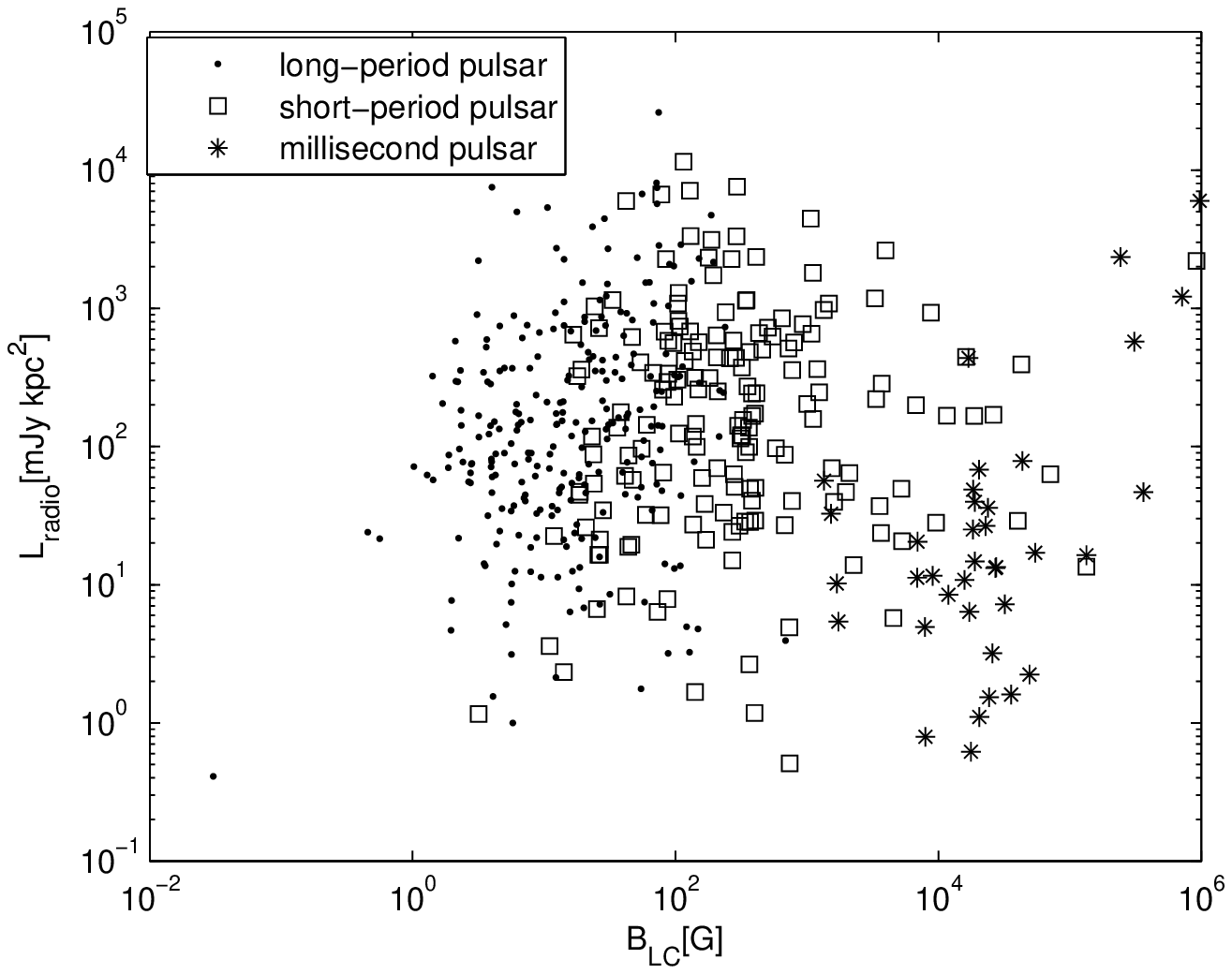}}
    \caption{Radio emission properties of normal and millisecond pulsars vs. fundamental parameters; \textit{upper panel}: the radio efficiency, $L_{\rm radio}/\dot{E}$, as function of the rate of rotational energy loss; \textit{lower panel}: the radio luminosity as function of the light-cylinder magnetic field. The data are taken from the ATNF Pulsar Catalogue}
\label{f4}
\end{figure}

As can be drawn from the two panels in Fig.~\ref{f4}, the radio luminosity of the millisecond pulsars is an increasing function of both $\dot{E}$ and $B_{\rm LC}$, in contrast to the other radio pulsars. One of the consequences is that weak radio millisecond pulsars are probably dim in gamma-rays as well, since the gamma-ray luminosity is known to increase with $\dot{E}$ \citep[e.g.,][]{fc2}. Then it is hard to discover such objects in any range. Perhaps, radio dim and radio quiet millisecond pulsars are yet to be detected.

A similar conclusion was made quite recently in \citet{calore15} based on an analysis of the direct dependency between the radio and gamma-ray luminosities of the millisecond pulsars, which was called the radio/gamma-ray correlation. (Note that in this case the term 'correlation' concerns a joint change of the average radio and gamma-ray luminosities from pulsar to pulsar rather than simultaneous temporal fluctuations of radiation in the two ranges, as is discussed in Sect.~2.3 for the Crab pulsar.) The authors do not discuss its physical nature, and our speculations are given below.

Besides a hint at the radio/gamma-ray correlation, Fig.~\ref{f4} raises one more interesting question: why the millisecond pulsars exhibit so much different behaviour of the radio luminosity as compared to the normal radio pulsars. The radio observations provide bulk of evidence that the radio properties of the normal and millisecond pulsars do not differ dramatically \citep[e.g.,][]{wiel98}, strongly testifying to the action of the same radio emission mechanism. We suggest that propagation effects control the radio luminosities of the millisecond pulsars.

As is already mentioned in Sect.~3.1, higher values of $B_{\rm LC}$ imply that the resonance region of radio waves lies higher in the magnetosphere. In the case of millisecond pulsars, the rotational effect is so strong that the radio beam may deflect into the outer/slot gap region even before the resonance. Then the radio beam interacts with the plasma flow, which is also a source of high-energy emission, and, in particular, may suffer induced scattering in different regimes, starting with the strong-field scattering.

The induced scattering of radio beam in different regimes may account for the additional components in the radio profiles of the millisecond pulsars. Because of concurrent scattering in the two polarization channels, the scattered emission may have low linear polarization. As the radio beam scattering takes place in the narrow regions of the outer magnetosphere, the position angle of linear polarization should be almost constant across the scattered components. We suggest a flat position angle curve as a definitive signature of the scattered components, as opposed to the S-shaped curve characteristic of the main pulse originating in the inner magnetosphere. The properties of the scattered components are in a general agreement with those observed \citep[for a comprehensive survey of the radio polarization profiles of millisecond pulsars see][]{dai15}.

Note that the overall energetics of the radio profile containing the scattered components may be higher, on account of the energy taken from the scattering particles and because of intensity transfer between substantially different spectral regions. The light-cylinder magnetic field predetermines the locus of the scattering regions, whereas the rate of rotational energy loss controlls the energetics of the scattering particles, so that the resultant radio luminosity is indeed expected to depend on both $B_{\rm LC}$ and $\dot{E}$.

Within the framework of our considerations, the radio/gamma-ray correlation in the millisecond pulsars can be explained by the fact that the two types of emission result from the same plasma flow related to the outer/slot gap. Note that an idea of direct generation of radio emission in the high-altitude gaps \citep[e.g.,][]{espinoza13,bilous15} encounters a difficulty of providing the necessary coherence in the outer magnetosphere. Therefore the propagation origin of the radio components correlating with gamma-ray emission seems preferable.

The induced scattering may give rise to the radio components leading and lagging the gamma-ray peak as well as those shifted by about a half of period from the main pulse. In a particular case when the radio and gamma-ray components are aligned, they are generated not only by the same particles but also at the same place. The correlation of such a sort \citep[the coincidence of radio and gamma-ray components is termed radio/gamma-ray correlation, e.g., in][]{espinoza13} is believed to be most informative and can be a useful tool of the plasma diagnostics in the high-altitude gaps.

In the most studied case of PSR J1939+2134, which represents the classic example of the coincident radio and gamma-ray peaks, the primordial main radio pulse (exhibiting an S-shaped position angle swing and, correspondingly, originating in the inner magnetosphere) actually precedes the gamma-ray peak, appears strongly suppressed and does not even look as a separate component \citep[see, e.g.,][]{dai15}. It is the other component, with the flat position angle curve and low linear polarization, that is responsible for the radio peak in the profile and coincides with the gamma-ray component. Recall that the radio interpulse of this pulsar is also strongly suggested to result from the backward scattering (see Sect.~2.1.3). Furthermore, it is the coincident radio components that show giant pulse phenomenon which can also be attributed to the induced scattering (see Sect.~2.2.1).

In view of the above considerations, it becomes clear why the Crab pulsar, unique among the young pulsars, bears close resemblance to the gamma-ray millisecond pulsars. Its light-cylinder magnetic field is similar to that of the millisecond pulsars, and the period is short enough for the strong-field induced scattering of radio beam to take place in the region of the high-altitude magnetospheric gaps. The scattering nature of the Crab's radio components was discussed in Sect.~2.2.3.

\section{Outlook}
The propagation concept reviewed above proves to be fruitful for pulsar studies. It not only allows to explain and sometimes to predict the phenomena observed in pulsar radio emission but also gives a unique opportunity to probe the pulsar magnetospheric plasma. For long years, the propagation studies developed independently of the poorly known magnetospheric picture and radio emission mechanism, being practically the only means intended to link the radio emission theory and observations.

The observational progress of the last decade have changed this situation drastically. The non-stationary phenomena in pulsar radio emission, such as moding and nulling \citep{null07}, rotating radio transients \citep{rrat06} and pulsar intermittency \citep{intermit06}, which are in the focus of current research, challenge the classic theory of the pulsar magnetosphere, especially in view of the relation found between the radio emission properties and the pulsar slowdown \citep{lyne10}. The gamma-ray studies of pulsars also require an advanced magnetospheric theory.

The last decade is marked with a substantial progress in modelling the pulsar force-free magnetosphere \citep[e.g.,][]{spitk06,3d13,abin15}. The force-free theory will hopefully be a proper basis for ultimate clarifying the pulsar emission mechanisms. In this context, the radio propagation theory has great perspectives as to interpretation of observations and diagnostics of the force-free magnetosphere.

The author is grateful to the anonymous referees for their constructive comments and suggestions.
The paper have used the data from the ATNF Pulsar Catalogue, the EPN Database of Pulsar Radio Profiles, and the LAT Second Pulsar Catalog.

\bibliographystyle{jpp}


\begin{thebibliography}

\bibitem[Abdo et al.(2010)]{fc1}
Abdo, A. A., Ackermann, M., Ajello, M., Atwood, 
W. B., Axelsson, M., et al. 2010 The First Fermi Large Area Telescope 
Catalog of Gamma-ray Pulsars. {\it Astrophys. J. Suppl. Ser.} {\bf 187}, 460-494.

\bibitem[Abdo et al.(2013)]{fc2}
Abdo, A. A., Ajello, M., Allafort, A., Baldini, 
L., Ballet, J., et al. 2013 The Second Fermi Large Area Telescope Catalog 
of Gamma-Ray Pulsars. {\it Astrophys. J. Suppl. Ser.} {\bf 208}, 17. 

\bibitem[Aliu et al.(2012)]{aliu12}
Aliu, E., Archambault, S., Arlen, T., Aune, 
T., Beilicke, M., et al. 2012 Search for a Correlation between 
Very-high-energy Gamma Rays and Giant Radio Pulses in the Crab Pulsar. {\it 
The Astrophysical Journal} {\bf 760}, 136.

\bibitem[Arendt \& Eilek(2002)]{ae02}
Arendt, P. N., Jr., \& Eilek, J. A. 2002 Pair Creation in the Pulsar Magnetosphere. {\it The Astrophysical Journal} {\bf 581}, 451-469. 

\bibitem[Arons \& Barnard(1986)]{ab86}
Arons, J., \& Barnard, J. J. 1986 Wave propagation in pulsar magnetospheres - Dispersion relations and normal modes of plasmas in superstrong magnetic fields. {\it The Astrophysical Journal} {\bf 302}, 120-137.


\bibitem[Asseo, Pelletier, \& Sol(1990)]{a90}
Asseo, E., Pelletier, G., \& Sol, H. 1990 A non-linear radio pulsar emission mechanism. {\it Monthly Notices of the Royal Astronomical Society} {\bf 247}, 529-548.

\bibitem[Asseo \& Melikidze(1998)]{am98}
Asseo, E., \& Melikidze, G. I. 1998 Non-stationary pair plasma in a pulsar magnetosphere and the two-stream Instability. {\it Monthly Notices of the Royal Astronomical Society} {\bf 301}, 59-71. 

\bibitem[Barnard(1986)]{b86}
Barnard, J. J. 1986 Probing the magnetic field of radio pulsars - A reexamination of polarization position angle swings. {\it The Astrophysical Journal} {\bf 303}, 280-291. 

\bibitem[Beskin \& Philippov(2012)]{bp12}
Beskin, V. S., \& Philippov, A. A. 2012 On the mean profiles of radio pulsars - I. Theory of propagation effects. {\it Monthly Notices of the Royal Astronomical Society} {\bf 425}, 814-840.

\bibitem[Beskin \& Rafikov (2000)]{br00}
Beskin, V. S., \& Rafikov, R. R. 2000 On the particle acceleration near the light surface of radio pulsars. {\it Monthly Notices of the Royal Astronomical Society} {\bf 313}, 433-444. 

\bibitem[Bilous et al.(2011)]{bilous11}
Bilous, A. V., Kondratiev, V. I., 
McLaughlin, M. A., Ransom, S. M., Lyutikov, M., Mickaliger, M., 
\& Langston, G. I. 2011 Correlation of Fermi Photons with High-frequency Radio Giant Pulses from the Crab Pulsar. {\it The Astrophysical Journal} {\bf 728}, 110.

\bibitem[Bilous et al.(2012)]{bilous12}
Bilous, A. V., McLaughlin, M. A., 
Kondratiev, V. I., 
\& Ransom, S. M. 2012 Correlation of Chandra Photons with the Radio Giant Pulses from the Crab Pulsar. {\it The Astrophysical Journal} {\bf 749}, 24. 

\bibitem[Bilous et al.(2015)]{bilous15}
Bilous, A. V., Pennucci, T. T., Demorest, P., 
\& Ransom, S. M. 2015 A Broadband Radio Study of the Average Profile and Giant Pulses from PSR B1821-24A. {\it The Astrophysical Journal} {\bf 803}, 83.

\bibitem[Blandford \& Scharlemann(1976)]{bs76}
Blandford, R. D., \& Scharlemann, E. T. 1976 On the scattering and absorption of electromagnetic radiation within pulsar magnetospheres. {\it Monthly Notices of the Royal Astronomical Society} {\bf 174}, 59-85. 

\bibitem[Brambilla et al.(2015)]{b15}
Brambilla, G., Kalapotharakos, C., 
Harding, A. K., 
\& Kazanas, D. 2015 Testing Dissipative Magnetosphere Model Light Curves and Spectra with Fermi Pulsars. {\it The Astrophysical Journal} {\bf 804}, 84. 

\bibitem[Cairns(2004)]{c04}
Cairns, I. H. 2004 Properties and Interpretations 
of Giant Micropulses and Giant Pulses from Pulsars. {\it The Astrophysical 
Journal} {\bf 610}, 948-955. 

\bibitem[Calore et al.(2015)]{calore15}
Calore, F., Di Mauro, M., Donato, F., Hessels, J. W. T., 
\& Weniger, C. 2015 Radio detection prospects for a bulge population of millisecond pulsars as suggested by Fermi LAT observations of the inner Galaxy. {\it ArXiv e-prints} arXiv:1512.06825.

\bibitem[Canuto, Lodenquai, \& Ruderman(1971)]{c71}
Canuto, V., Lodenquai, J., \& Ruderman, M. 1971 Thomson Scattering in a Strong Magnetic Field. {\it Physical Review D} {\bf 3}, 2303-2308. 

\bibitem[Cerutti, Philippov, 
\& Spitkovsky(2016)]{spitk15}
Cerutti, B., Philippov, A. A., \& Spitkovsky, A. 2016 Modeling high-energy pulsar lightcurves from first principles. {\it Monthly Notices of the Royal 
Astronomical Society} {\bf 457}, 2401-2414.

\bibitem[Cheng \& Ruderman(1979)]{cr79}
Cheng, A. F., \& Ruderman, M. A. 1979 A theory of subpulse polarization patterns from radio pulsars. {\it The Astrophysical Journal} {\bf 229}, 348-360. 

\bibitem[Cheng, Ruderman, 
\& Zhang(2000)]{cheng00}
Cheng, K. S., Ruderman, M., \& Zhang, L. 2000 A Three-dimensional Outer Magnetospheric Gap Model for Gamma-Ray Pulsars: Geometry, Pair Production, Emission Morphologies, and Phase-resolved Spectra. {\it The Astrophysical Journal} {\bf 537}, 964-976.

\bibitem[Dai et al.(2015)]{dai15}
Dai, S., Hobbs, G., Manchester, R. N., Kerr, 
M., Shannon, R. M., et al. 2015 A study of multifrequency polarization 
pulse profiles of millisecond pulsars. {\it Monthly Notices of the Royal 
Astronomical Society} {\bf 449}, 3223-3262. 

\bibitem[Dyks, Zhang, 
\& Gil(2005)]{d05} Dyks, J., Zhang, B., \& Gil, J. 2005 Reversals of Radio Emission Direction in PSR B1822-09. {\it The Astrophysical Journal} {\bf 626}, L45-L47. 

\bibitem[Edwards(2004)]{e04}
Edwards, R. T. 2004 The polarization of drifting 
subpulses. {\it Astronomy and Astrophysics} {\bf 426}, 677-686. 

\bibitem[Edwards \& Stappers(2004)]{es04}
Edwards, R. T., \& Stappers, B. W. 2004 Ellipticity and deviations from orthogonality in the polarization modes of PSR B0329+54. {\it Astronomy and Astrophysics} {\bf 421}, 681-691.

\bibitem[Espinoza et al.(2013)]{espinoza13}
Espinoza, C. M., Guillemot, L., Celik, O., Weltevrede, P., Stappers, B. W., et al. 2013 Six millisecond pulsars detected by the Fermi Large Area Telescope and the radio/gamma-ray connection of millisecond pulsars. {\it Monthly Notices of the Royal Astronomical Society} {\bf 430}, 571-587.

\bibitem[Fowler, Morris, 
\& Wright (1981)]{f81}
Fowler, L. A., Morris, D., \& Wright, G. A. E. 1981 Unusual properties of the pulsar PSR 1822-09. {\it Astronomy and Astrophysics} {\bf 93}, 54-61. 

\bibitem[Gangadhara \& Krishan(1993)]{gang93}
Gangadhara, R. T., \& Krishan, V. 1993 Polarization Changes of Radiation through Stimulated Raman Scattering. {\it The Astrophysical Journal} {\bf 415}, 505. 

\bibitem[Gil et al.(1994)]{g94}
Gil, J. A., Jessner, A., Kijak, J., Kramer, 
M., Malofeev, V., et al. 1994 Multifrequency study of PSR 1822-09. {\it 
Astronomy and Astrophysics} {\bf 282}, 45-53. 

\bibitem[Gil, Khechinashvili, 
\& Melikidze(2001)]{g01}
Gil, J. A., Khechinashvili, D. G., \& Melikidze, G. I. 2001 On the Optical Pulsations from the Geminga Pulsar. {\it The Astrophysical Journal} {\bf 551}, 867-873.

\bibitem[Goldreich \& Julian(1969)]{gj69}
Goldreich, P., \& Julian, W. H. 1969 Pulsar ectrodynamics. {\it The Astrophysical Journal} {\bf 157}, 869.

\bibitem[Hankins et al.(2003)]{h03}
Hankins, T. H., Kern, J. S., Weatherall, 
J. C., \& Eilek, J. A. 2003 Nanosecond radio bursts from strong plasma turbulence in the Crab pulsar. {\it Nature} {\bf 422}, 141-143.

\bibitem[Hankins 
\& Eilek(2007)]{he07}
Hankins, T. H., \& Eilek, J. A. 2007 Radio Emission Signatures in the Crab Pulsar. {\it The Astrophysical Journal} {\bf 670}, 693-701. 

\bibitem[Hankins, Jones, 
\& Eilek(2015)]{h15}
Hankins, T. H., Jones, G., \& Eilek, J. A. 2015 The Crab Pulsar at Centimeter Wavelengths. I. Ensemble Characteristics. {\it The Astrophysical Journal} {\bf 802}, id.130, 12pp. 

\bibitem[Harding et al.(2002)]{hard02}
Harding, A. K., Strickman, M. S., Gwinn, 
C., Dodson, R., Moffet, D., 
\& McCulloch, P. 2002 The Multicomponent Nature of the Vela Pulsar Nonthermal X-Ray Spectrum. {\it The Astrophysical Journal} {\bf 576}, 376-380. 

\bibitem[Harding, Usov, 
\& Muslimov(2005)]{hum05}
Harding, A. K., Usov, V. V., \& Muslimov, A. G. 2005 High-Energy Emission from Millisecond Pulsars. {\it The Astrophysical Journal} {\bf 622}, 531-543.

\bibitem[Harding et al.(2008)]{hard08}
Harding, A. K., Stern, J. V., Dyks, J., 
\& Frackowiak, M. 2008 High-Altitude Emission from Pulsar Slot Gaps: The Crab Pulsar. {\it The Astrophysical Journal} {\bf 680}, 1378-1393.

\bibitem[Harding \& Kalapotharakos (2015)]{hk15}
Harding, A. K., \& Kalapotharakos, C. 2015 Synchrotron Self-Compton Emission from the Crab and Other Pulsars. {\it The Astrophysical Journal} {\bf 811}, 63. 

\bibitem[Hibschman \& Arons(2001a)]{ha01a}
Hibschman, J. A., \& Arons, J. 2001a Polarization Sweeps in Rotation-powered Pulsars. {\it The Astrophysical Journal} {\bf 546}, 382-393.

\bibitem[Hibschman \& Arons(2001b)]{ha01b}
Hibschman, J. A., \& Arons, J. 2001b Pair Production Multiplicities in Rotation-powered Pulsars. {\it The Astrophysical Journal} {\bf 560}, 871-884. 

\bibitem[Hirotani(2006)]{hirot06}
Hirotani, K. 2006 High-Energy Emission from 
Pulsar Magnetospheres. {\it Modern Physics Letters A} {\bf 21}, 1319-1337. 

\bibitem[Ichimaru(1970)]{i70}
Ichimaru, S. 1970 Plasma Turbulence as a 
Mechanism of Pulsar Radiation. {\it Nature} {\bf 226}, 731-733.

\bibitem[Johnson et al.(2014)]{v14}
Johnson, T. J., Venter, C., Harding, A. 
K., Guillemot, L., Smith, D. A., et al. 2014 Constraints on the Emission 
Geometries and Spin Evolution of Gamma-Ray Millisecond Pulsars. {\it The 
Astrophysical Journal Supplement Series} {\bf 213}, 6, 1-54.

\bibitem[Johnston, Karastergiou, \& Willett(2006)]{vela2}
Johnston, S., Karastergiou, A., \& Willett, K. 2006 High-frequency observations of southern pulsars. {\it Monthly Notices of the Royal Astronomical Society} {\bf 369}, 1916-1928.

\bibitem[Kalapotharakos et al.(2012)]{k12} Kalapotharakos, C., Kazanas, D., 
Harding, A., 
\& Contopoulos, I. 2012 Toward a Realistic Pulsar Magnetosphere. {\it The Astrophysical Journal} {\bf 749}, 2. 

\bibitem[Kalapotharakos, Harding, 
\& Kazanas (2014)]{k14}
Kalapotharakos, C., Harding, A. K., \& Kazanas, D. 2014 Gamma-Ray Emission in Dissipative Pulsar Magnetospheres: From Theory to Fermi Observations. {\it The Astrophysical Journal} {\bf 793}, 97. 

\bibitem[Karastergiou \& Johnston(2006)]{vela1}
Karastergiou, A., \& Johnston, S. 2006 Absolute polarization position angle profiles of southern pulsars at 1.4 and 3.1 GHz. {\it Monthly Notices of the Royal Astronomical Society} {\bf 365}, 353-366.

\bibitem[Knight et al.(2006a)]{k06a}
Knight, H. S., Bailes, M., Manchester, R. N., Ord, S. M., 
\& Jacoby, B. A. 2006a Green Bank Telescope Studies of Giant Pulses from Millisecond Pulsars. {\it The Astrophysical Journal} {\bf 640}, 941-949. 

\bibitem[Knight et al.(2006b)]{k06b}
Knight, H. S., Bailes, M., Manchester, R. N., 
\& Ord, S. M. 2006b A Study of Giant Pulses from PSR J1824-2452A. {\it The Astrophysical Journal} {\bf 653}, 580-586. 

\bibitem[Kojima \& Oogi (2009)]{ko09}
Kojima, Y., \& Oogi, J. 2009 Numerical construction of magnetosphere with relativistic two-fluid plasma flows. {\it Monthly Notices of the Royal Astronomical Society} {\bf 398}, 271-279. 

\bibitem[Kramer et al.(1998)]{wiel98}
Kramer, M., Xilouris, K. M., Lorimer, D. 
R., Doroshenko, O., Jessner, A., Wielebinski, R., Wolszczan, A., 
\& Camilo, F. 1998 The Characteristics of Millisecond Pulsar Emission. I. Spectra, Pulse Shapes, and the Beaming Fraction. {\it The Astrophysical Journal} {\bf 501}, 270-285.

\bibitem[Kramer et al.(2006)]{intermit06}
Kramer, M., Lyne, A. G., O'Brien, J. T., Jordan, C. A., 
\& Lorimer, D. R. 2006 A Periodically Active Pulsar Giving Insight into Magnetospheric Physics. {\it Science} {\bf 312}, 549-551. 

\bibitem[Krishnamohan 
\& Downs(1983)]{kd83}
Krishnamohan, S., \& Downs, G. S. 1983 Intensity dependence of the pulse profile and polarization of the VELA pulsar. {\it The Astrophysical Journal} {\bf 265}, 372-388.

\bibitem[Li, Spitkovsky, 
\& Tchekhovskoy(2012)]{lst12}
Li, J., Spitkovsky, A., \& Tchekhovskoy, A. 2012 Resistive Solutions for Pulsar Magnetospheres. {\it The Astrophysical Journal} {\bf 746}, 60. 

\bibitem[Lommen et al.(2007)]{lommen07}
Lommen, A., Donovan, J., Gwinn, C., 
Arzoumanian, Z., Harding, A., et al. 2007 Correlation between X-Ray 
Light-Curve Shape and Radio Arrival Time in the Vela Pulsar. {\it The 
Astrophysical Journal} {\bf 657}, 436-440.

\bibitem[Lundgren et al.(1995)]{lund95}
Lundgren, S. C., Cordes, J. M., Ulmer, 
M., Matz, S. M., Lomatch, S., Foster, R. S., 
\& Hankins, T. 1995 Giant Pulses from the Crab Pulsar: A Joint Radio and Gamma-Ray Study. {\it The Astrophysical Journal} {\bf 453}, 433-445.

\bibitem[Luo \& Melrose(2006)]{lm06}
Luo, Q., \& Melrose, D. B. 2006 The induced turbulence effect on propagation of radio emission in pulsar magnetospheres. {\it Monthly Notices of the Royal Astronomical Society} {\bf 371}, 1395-1404.

\bibitem[Lyne et al.(2010)]{lyne10}
Lyne, A., Hobbs, G., Kramer, M., Stairs, I., 
\& Stappers, B. 2010 Switched Magnetospheric Regulation of Pulsar Spin-Down. {\it Science} {\bf 329}, 408-412.

\bibitem[Lyubarsky(1995)]{l95}
Lyubarsky, Y. E. 1995 Physics of pulsars. {\it Amsterdam: Harwood Academic Publishers}. 

\bibitem[Lyubarskii(1996)]{l96} Lyubarskii, Y. E. 1996 Generation of pulsar 
radio emission. {\it Astronomy and Astrophysics} {\bf 308}, 809-820. 

\bibitem[Lyubarsky(2008)]{l08}
Lyubarsky, Y. 2008 Pulsar emission mechanisms. {\it 40 Years of Pulsars: Millisecond Pulsars, Magnetars and More} {\bf 983}, 29-37. 

\bibitem[Lyubarskii \& Petrova (1996)]{lp96}
Lyubarskii, Y. E., \& Petrova, S. A. 1996 Stimulated scattering of radio emission in pulsar magnetospheres. {\it Astronomy Letters} {\bf 22}, 399-408.

\bibitem[Lyubarskii \& Petrova(1998a)]{lp98a}
Lyubarskii, Y. E., \& Petrova, S. A. 1998a On the Circular Polarization of Pulsar Radiation. {\it Astrophysics and Space Science} {\bf 262}, 379-389.  

\bibitem[Lyubarskii \& Petrova(1998b)]{lp98b}
Lyubarskii, Y. E., \& Petrova, S. A. 1998 Synchrotron absorption in pulsar magnetospheres. {\it Astronomy and Astrophysics} {\bf 337}, 433-440. 

\bibitem[Lyubarskii \& Petrova(2000)]{lp00}
Lyubarskii, Y. E., \& Petrova, S. A. 2000 Resonant inverse Compton scattering by secondary pulsar plasma. {\it Astronomy and Astrophysics} {\bf 355}, 406-412.

\bibitem[Lyutikov(1998a)]{lyut98a}
Lyutikov, M. 1998a Waves in a one-dimensional magnetized relativistic pair plasma. {\it Monthly Notices of the Royal Astronomical Society} {\bf 293}, 447.

\bibitem[Lyutikov(1998b)]{lyut98b}
Lyutikov, M. 1998b Induced Raman scattering in pulsar magnetospheres. {\it Monthly Notices of the Royal Astronomical Society} {\bf 298}, 1198-1206. 


\bibitem[Lyutikov, Otte, 
\& McCann(2012)]{lyut12}
Lyutikov, M., Otte, N., \& McCann, A. 2012 The Very High Energy Emission from Pulsars: A Case for Inverse Compton Scattering. {\it The Astrophysical Journal} {\bf 754}, 33.

\bibitem[Lyutikov(2013)]{lyut13}
Lyutikov, M. 2013 Inverse Compton model of 
pulsar high-energy emission. {\it Monthly Notices of the Royal Astronomical 
Society} {\bf 431}, 2580-2589.

\bibitem[Marelli et al.(2015)]{marelli15}
Marelli, M., Mignani, R. P., De Luca, 
A., Saz Parkinson, P. M., Salvetti, D., Den Hartog, P. R., 
\& Wolff, M. T. 2015 Radio-quiet and Radio-loud Pulsars: Similar in Gamma-Rays but Different in X-Rays. {\it The Astrophysical Journal} {\bf 802}, 78.

\bibitem[McKinnon 
\& Stinebring(2000)]{ms00}
McKinnon, M. M., \& Stinebring, D. R. 2000 The Mode-separated Pulse Profiles of Pulsar Radio Emission. {\it The Astrophysical Journal} {\bf 529}, 435-446.

\bibitem[McLaughlin et al.(2006)]{rrat06}
McLaughlin, M. A., Lyne, A. G., 
Lorimer, D. R., Kramer, M., Faulkner, A. J., et al. 2006 Transient radio 
bursts from rotating neutron stars. {\it Nature} {\bf 439}, 817-820.

\bibitem[Medin \& Lai(2010)]{ml10}
Medin, Z., \& Lai, D. 2010 Pair cascades in the magnetospheres of strongly magnetized neutron stars. {\it Monthly Notices of the Royal Astronomical Society} {\bf 406}, 1379-1404.

\bibitem[Melrose(2003)]{m03}
Melrose, D. B. 2003 Pulsar emissions. {\it Plasma Physics and Controlled Fusion} {\bf 45}, 523-534.

\bibitem[Melrose \& Stoneham(1977)]{ms77}
Melrose, D. B., \& Stoneham, R. J. 1977 The natural wave modes in a pulsar magnetosphere. {\it Proceedings of the Astronomical Society of Australia} {\bf 3}, 120-122. 

\bibitem[Melrose \& Gedalin(1999)]{mg99}
Melrose, D. B., \& Gedalin, M. E. 1999 Relativistic Plasma Emission and Pulsar Radio Emission: A Critique. {\it The Astrophysical Journal} {\bf 521}, 351-361.

\bibitem[Melrose et al.(1999)]{m99}
Melrose, D. B., Gedalin, M. E., Kennett, M. P., 
\& Fletcher, C. S. 1999 Dispersion in an intrinsically relativistic, one-dimensional, strongly magnetized pair plasma. {\it Journal of Plasma Physics} {\bf 62}, 233-248. 

\bibitem[Mikhailovskii et al.(1982)]{m82}
Mikhailovskii, A. B., Onishchenko, 
O. G., Suramlishvili, G. I., 
\& Sharapov, S. E. 1982 The Emergence of Electromagnetic Waves from Pulsar Magnetospheres. {\it Soviet Astronomy Letters} {\bf 8}, 369-371. 

\bibitem[Mochol 
\& P{\'e}tri (2015)]{mp15}
Mochol, I., \& P{\'e}tri, J. 2015 Very high energy emission as a probe of relativistic magnetic reconnection in pulsar winds. {\it Monthly Notices of the Royal Astronomical Society} {\bf 449}, L51-L55.

\bibitem[Moffett 
\& Hankins(1996)]{mh96} Moffett, D. A., \& Hankins, T. H. 1996 Multifrequency Radio Observations of the Crab Pulsar. {\it The Astrophysical Journal} {\bf 468},779. 

\bibitem[Moffett 
\& Hankins(1999)]{mh99} Moffett, D. A., \& Hankins, T. H. 1999 Polarimetric Properties of the Crab Pulsar between 1.4 and 8.4 GHZ. {\it The Astrophysical Journal} {\bf 522}, 1046-1052. 

\bibitem[Muslimov 
\& Harding(2004)]{mh04}
Muslimov, A. G., \& Harding, A. K. 2004 High-Altitude Particle Acceleration and Radiation in Pulsar Slot Gaps. {\it The Astrophysical Journal} {\bf 606}, 1143-1153.

\bibitem[Ochelkov 
\& Usov (1983)]{ochusov83}
Ochelkov, I. P., \& Usov, V. V. 1983 Compton scattering of electromagnetic radiation in pulsar magnetospheres. {\it Astrophysics and Space Science} {\bf 96}, 55-81. 

\bibitem[Oosterbroek et al.(2008)]{oosterbr08}
Oosterbroek, T., Cognard, I., 
Golden, A., Verhoeve, P., Martin, D. D. E., et al. 2008 Simultaneous 
absolute timing of the Crab pulsar at radio and optical wavelengths. {\it 
Astronomy and Astrophysics} {\bf 488}, 271-277.

\bibitem[P{\'e}tri(2012)]{petri12}
P{\'e}tri, J. 2012 High-energy emission from the 
pulsar striped wind: a synchrotron model for gamma-ray pulsars. {\it 
Monthly Notices of the Royal Astronomical Society} {\bf 424}, 2023-2027.

\bibitem[Petrova \& Lyubarskii(2000)]{pl00}
Petrova, S. A., \& Lyubarskii, Y. E. 2000 Propagation effects in pulsar magnetospheres. {\it Astronomy and Astrophysics} {\bf 355}, 1168-1180.

\bibitem[Petrova(2001)]{p01}
Petrova, S. A. 2001 On the origin of orthogonal 
polarization modes in pulsar radio emission. {\it Astronomy and 
Astrophysics} {\bf 378}, 883-897.

\bibitem[Petrova(2002)]{p02}
Petrova, S. A. 2002 The effect of synchrotron 
absorption on the observed radio luminosities of pulsars. {\it Monthly 
Notices of the Royal Astronomical Society} {\bf 336}, 774-784.

\bibitem[Petrova(2003a)]{p03a}
Petrova, S. A. 2003a Diagnostics of the plasma of 
pulsar magnetospheres based on polarization profiles of radio pulses. {\it 
Astronomy and Astrophysics} {\bf 408}, 1057-1063.

\bibitem[Petrova(2003b)]{p03b}
Petrova, S. A. 2003b A model for non-thermal 
optical emission of radio pulsars. {\it Monthly Notices of the Royal 
Astronomical Society} {\bf 340}, 1229-1239.

\bibitem[Petrova(2004a)]{p04a}
Petrova, S. A. 2004a Toward explanation of 
microstructure in pulsar radio emission. {\it Astronomy and Astrophysics} 
{\bf 417}, L29-L32.

\bibitem[Petrova(2004b)]{p04b}
Petrova, S. A. 2004b On the origin of giant 
pulses in radio pulsars. {\it Astronomy and Astrophysics} {\bf 424}, 
227-236.

\bibitem[Petrova(2006a)]{p06a}
Petrova, S. A. 2006a Polarization transfer in a 
pulsar magnetosphere. {\it Monthly Notices of the Royal Astronomical 
Society} {\bf 366}, 1539-1550.

\bibitem[Petrova(2006b)]{p06b}
Petrova, S. A. 2006b Statistics of the 
individual-pulse polarization based on propagation effects in the pulsar 
magnetosphere. {\it Monthly Notices of the Royal Astronomical Society} {\bf 
368}, 1764-1772.

\bibitem[Petrova(2008a)]{p08a}
Petrova, S. A. 2008a Physics of Interpulse 
Emission in Radio Pulsars. {\it The Astrophysical Journal} {\bf 673}, 
400-410.

\bibitem[Petrova(2008b)]{p08b}
Petrova, S. A. 2008b Scattering of low-frequency 
radiation by a gyrating electron. {\it Monthly Notices of the Royal 
Astronomical Society} {\bf 383}, 1413-1424.

\bibitem[Petrova(2008c)]{p08c}
Petrova, S. A. 2008c On the nature of precursors 
in the radio pulsar profiles. {\it Monthly Notices of the Royal 
Astronomical Society} {\bf 384}, L1-L5.

\bibitem[Petrova (2008d)]{p08d}
Petrova, S. A. 2008d Interpretation of the 
low-frequency peculiarities in the radio profile structure of the Crab 
pulsar. {\it Monthly Notices of the Royal Astronomical Society} {\bf 385}, 
2143-2150.

\bibitem[Petrova(2009a)]{p09a}
Petrova, S. A. 2009a Towards explanation of the 
X-ray-radio correlation in the Vela pulsar. {\it Monthly Notices of the 
Royal Astronomical Society} {\bf 395}, 290-300.

\bibitem[Petrova(2009b)]{p09b}
Petrova, S. A. 2009b Formation of the radio 
profile components of the Crab pulsar. {\it Monthly Notices of the Royal 
Astronomical Society} {\bf 395}, 1723-1732. 

\bibitem[Petrova(2015)]{p15}
Petrova, S. A. 2015 Axisymmetric force-free 
magnetosphere of a pulsar - II. Transition from the self-consistent 
two-fluid model. {\it Monthly Notices of the Royal Astronomical Society} 
{\bf 446}, 2243-2250. 

\bibitem[Philippov, Spitkovsky, 
\& Cerutti(2015)]{abin15}
Philippov, A. A., Spitkovsky, A., \& Cerutti, B. 2015 Ab Initio Pulsar Magnetosphere: Three-dimensional Particle-in-cell Simulations of Oblique Pulsars. {\it The Astrophysical Journal} {\bf 801}, L19.

\bibitem[Pierbattista et al.(2015)]{pier15}
Pierbattista, M., Harding, A. K., 
Grenier, I. A., Johnson, T. J., Caraveo, P. A., Kerr, M., 
\& Gonthier, P. L. 2015 Light-curve modelling constraints on the obliquities and aspect angles of the young Fermi pulsars. {\it Astronomy and Astrophysics} {\bf 575}, A3. 

\bibitem[Radhakrishnan \& Cooke(1969)]{rc69}
Radhakrishnan, V., \& Cooke, D. J. 1969 Magnetic Poles and the Polarization Structure of Pulsar Radiation. {\it Astrophysical Letters} {\bf 3}, 225. 

\bibitem[Ramanamurthy 
\& Thompson (1998)]{rt98}
Ramanamurthy, P. V., \& Thompson, D. J. 1998 Search for Short-Term Variations in the 50 MeV Gamma-Ray Emission of the Crab Pulsar. {\it The Astrophysical Journal} {\bf 496}, 863-868. 

\bibitem[Romani(1996)]{romani96}
Romani, R. W. 1996 Gamma-Ray Pulsars: Radiation 
Processes in the Outer Magnetosphere. {\it The Astrophysical Journal} {\bf 
470}, 469.

\bibitem[Shearer et al.(2003)]{shearer03}
Shearer, A., Stappers, B., O'Connor, P., 
Golden, A., Strom, R., Redfern, M., 
\& Ryan, O. 2003 Enhanced Optical Emission During Crab Giant Radio Pulses. {\it Science} {\bf 301}, 493-495. 

\bibitem[Spitkovsky(2006)]{spitk06}
Spitkovsky, A. 2006 Time-dependent Force-free 
Pulsar Magnetospheres: Axisymmetric and Oblique Rotators. {\it The 
Astrophysical Journal} {\bf 648}, L51-L54.

\bibitem[Strader et al.(2013)]{s13}
Strader, M. J., Johnson, M. D., Mazin, 
B. A., Spiro Jaeger, G. V., Gwinn, C. R., et al. 2013 Excess Optical 
Enhancement Observed with ARCONS for Early Crab Giant Pulses. {\it The 
Astrophysical Journal} {\bf 779}, L12.

\bibitem[Sturrock(1970)]{s70}
Sturrock, P. A. 1970 Pulsar Radiation 
Mechanisms. {\it Nature} {\bf 227}, 465-470.

\bibitem[Takata, Wang, \& Cheng(2010)]{twc10}
Takata, J., Wang, Y., \& Cheng, K. S. 2010 Pulsar High Energy Emissions from Outer Gap Accelerator Closed by a Magnetic Pair-creation Process. {\it The Astrophysical Journal} {\bf 715}, 1318-1326.

\bibitem[Tanaka, Asano, 
\& Terasawa (2015)]{t15}
Tanaka, S. J., Asano, K., \& Terasawa, T. 2015 Avalanche photon cooling by induced Compton scattering: Higher-order Kompaneets equation. {\it Progress of Theoretical and Experimental Physics} {\bf 2015}, 073E01, 1-14.

\bibitem[Tanaka \& Takahara(2013)]{tt13}
Tanaka, S. J., \& Takahara, F. 2013 Constraint on Pulsar Wind Properties from Induced Compton Scattering off Radio Pulses. {\it Progress of Theoretical and Experimental Physics} {\bf 12}, 123E01. 

\bibitem[Tchekhovskoy, Spitkovsky, \& Li(2013)]{3d13}
Tchekhovskoy, A., Spitkovsky, A., \& Li, J. G. 2013 Time-dependent 3D magnetohydrodynamic pulsar magnetospheres: oblique rotators. {\it Monthly Notices of the Royal Astronomical Society} {\bf 435}, L1-L5.

\bibitem[Timokhin \& Harding (2015)]{timhar15}
Timokhin, A. N., \& Harding, A. K. 2015 On the Polar Cap Cascade Pair Multiplicity of Young Pulsars. {\it The Astrophysical Journal} {\bf 810}, 144.

\bibitem[Ursov \& Usov(1988)]{uu88}
Ursov, V. N., \& Usov, V. V. 1988 Plasma flow nonstationarity in pulsar magnetospheres and two-stream instability. {\it Astrophysics and Space Science} {\bf 140}, 325-336. 

\bibitem[Uzdensky 
\& Spitkovsky (2014)]{u14}
Uzdensky, D. A., \& Spitkovsky, A. 2014 Physical Conditions in the Reconnection Layer in Pulsar Magnetospheres. {\it The Astrophysical Journal} {\bf 780}, 3. 

\bibitem[Wang, Lai, \& Han(2010)]{wang10}
Wang, C., Lai, D., \& Han, J. 2010 Polarization changes of pulsars due to wave propagation through magnetospheres. {\it Monthly Notices of the Royal Astronomical Society} {\bf 403}, 569-588. 

\bibitem[Wang, Manchester, 
\& Johnston(2007)]{null07}
Wang, N., Manchester, R. N., \& Johnston, S. 2007 Pulsar nulling and mode changing. {\it Monthly Notices of the Royal Astronomical Society} {\bf 377}, 1383-1392.

\bibitem[Weatherall(1994)]{w94}
Weatherall, J. C. 1994 Streaming instability in relativistically hot pulsar magnetospheres. {\it The Astrophysical Journal} {\bf 428}, 261-266. 

\bibitem[Weatherall(1998)]{w98}
Weatherall, J. C. 1998 Pulsar Radio Emission 
by Conversion of Plasma Wave Turbulence: Nanosecond Time Structure. {\it 
The Astrophysical Journal} {\bf 506}, 341-346.

\bibitem[Weltevrede, Wright, 
\& Stappers(2007)]{welt07}
Weltevrede, P., Wright, G. A. E., \& Stappers, B. W. 2007 The main-interpulse interaction of PSR B1702-19. {\it Astronomy and Astrophysics} {\bf 467}, 1163-1174. 

\bibitem[Wilson \& Rees(1978)]{wr78}
Wilson, D. B., \& Rees, M. J. 1978 Induced Compton scattering in pulsar winds. {\it Monthly Notices of the Royal Astronomical Society} {\bf 185}, 297. 

\end{thebibliography}

\end{document}